\documentclass[a4paper,12pt]{article}
\pdfoutput=1

\usepackage{jheppub}
\usepackage[sort&compress]{natbib}
\usepackage[utf8]{inputenc}
\usepackage{amsmath,amssymb,amsthm}
\usepackage{physics,mathtools}
\usepackage{float}
\usepackage{subfig}
\usepackage[svgnames]{xcolor}

\usepackage{comment}
\usepackage[normalem]{ulem}

\hypersetup{linkcolor=DarkBlue,citecolor=DarkBlue, urlcolor=black} 
\makeatletter
\def\@fpheader{\vspace{1mm}}
\makeatother

\newcommand{\bw}{\begin{widetext}}
\newcommand{\ew}{\end{widetext}}
\newcommand{\bea}{\begin{eqnarray}}
\newcommand{\eea}{\end{eqnarray}}
\newcommand{\be}{\begin{equation}}
\newcommand{\ee}{\end{equation}}
\newcommand{\nn}{\nonumber}

\renewcommand{\bar}[1]{\overline{#1}}
\renewcommand{\tilde}[1]{\widetilde{#1}}
\renewcommand{\hat}[1]{\widehat{#1}}

\renewcommand{\abs}[1]{\left| #1 \right|}
\renewcommand{\cal}{\mathcal}

\newcommand{\ra}{\rightarrow}

\newcommand{\xp}{{x^+}}
\newcommand{\xpp}{{(x^+)}}
\newcommand{\xm}{{x^-}}

\newcommand{\qf}{\mathfrak{q}}
\newcommand{\wf}{\mathfrak{w}}

\newcommand{\ff}{{\mathfrak{f}}}
\newcommand{\Zt}{Z_{\rm scalar}} 
\newcommand{\Zv}{Z_{\rm  shear}}  
\newcommand{\Zs}{Z_{\rm  sound}}
\newcommand{\OO}{{\cal O}}

\renewcommand{\aa}{\hat{a}}
\newcommand{\bb}{\hat{b}}
\newcommand{\cc}{\hat{c}}

\def\qfr{\mathfrak{q}}
\def\wfr{\mathfrak{w}}
\def \nn {\nonumber}
 \def\dd {{\rm d}}
\def\wf{\mathfrak{w}}

\title{\LARGE{Thermal Stress Tensor Correlators near Lightcone and Holography}}  

\author{\vspace{-0.5cm}
Chantelle Esper,$^{\color{black}a}$}
\author{\hspace{0.02cm}Kuo-Wei Huang,$^{\color{black}a}$}
\author{\hspace{0.02cm}Robin Karlsson,$^{\color{black}b}$}
\author{\hspace{0.02cm}Andrei Parnachev,$^{\color{black}a}$\linebreak} 
\author{and Samuel Valach$^{\color{black}a}$\linebreak\vspace{-0.2cm}}

\affiliation[{\color{black}a}]{School of Mathematics and Hamilton Mathematics Institute, 
\setlength{\parskip}{0pt}\newline\indent Trinity College, Dublin 2, Ireland}
\affiliation[{\color{black}b}]{CERN, Theoretical Physics Department, CH-1211 Geneva 23, Switzerland}

\preprint{CERN-TH-2023-091}

\abstract{We consider thermal stress-tensor two-point functions in holographic theories in the near-lightcone regime and
analyse them using the operator product expansion (OPE). 
In the limit we consider only the leading-twist multi-stress tensors contribute and the correlators depend on a particular
combination of lightcone momenta.
We argue that such correlators are described by three universal functions, which can be holographically computed in Einstein 
gravity; higher-derivative terms in the gravitational Lagrangian enter the arguments of these functions
via the cubic stress-tensor couplings and the thermal stress-tensor expectation value in the dual CFT. 
We compute the retarded correlators and observe that in addition to the perturbative OPE, which contributes to the real part,
there is a non-perturbative contribution to the  imaginary part.

}

\begin{document}

\maketitle
\flushbottom
\newpage

\section{Introduction}
\label{sec.introduction}

Understanding the non-perturbative structure of quantum field theories (QFTs) at finite temperature is an important challenge of theoretical physics. 
In particular, thermal fluctuations cannot be ignored when studying real-world, out-of-equilibrium phenomena in, $e.g.$, the quark-gluon plasma or strongly coupled condensed matter systems.  
Conformal field theories (CFTs) provide a natural starting point for investigating QFTs more generally.  
The additional symmetries in conformal theories impose powerful constraints on physical observables, even at strong coupling. 
Achieving a better understanding of general thermal field theories should therefore begin with an investigation of CFTs at finite temperature. 

A basic probe in any local field theory is the stress tensor, $T_{\mu\nu}$. 
The importance of the stress tensor, and correlators thereof, to CFT is hard to overstate.  The stress-tensor sector is completely universal in two-dimensional CFT, where the infinite-dimensional Virasoro algebra allows determining physical observables based on symmetries.  In higher dimensions, although the structure of conformal correlators is less constrained and in general they are determined via case-by-case computations, the conformal invariance still fixes two- and three-point stress-tensor correlators uniquely up to a few constants \cite{Osborn:1993cr}.   In particular, the two-point stress-tensor correlator at zero temperature is fixed up to an overall coefficient, the central charge $C_T$. 

When considering the theory at finite temperature, the stress-tensor two-point functions are no longer fixed by one constant. Instead,  these thermal $TT$ correlators are generally theory dependent:  they depend on, for instance, the coefficients appearing in the zero-temperature three-point correlators of stress tensors. 
 It is desirable to identify physical limits that make some universal aspects of the thermal correlators  manifest, and then devote future efforts to computing non-universal corrections to the correlators.

In this paper, using the AdS/CFT correspondence \cite{Maldacena:1997re, Gubser:1998bc, Witten:1998qj}, we  analyze thermal $TT$ correlators at large central charge in a class of holographic CFTs  in four dimensions, focusing on a certain universal, near-lightcone regime.   
Our analysis of the near-lightcone $TT$ correlators  
is in part motivated by the large body of recent work on  thermal scalar correlators and their near-lightcone behavior in spacetime dimensions greater than two 
\cite{Kulaxizi:2018dxo,Fitzpatrick:2019zqz, Karlsson:2019qfi,Li:2019tpf,Huang:2019fog,Kulaxizi:2019tkd,Fitzpatrick:2019efk,Haehl:2019eae,Karlsson:2019dbd,Li:2019zba,Karlsson:2019txu,Huang:2020ycs,Karlsson:2020ghx,Li:2020dqm,Parnachev:2020fna,Fitzpatrick:2020yjb,Berenstein:2020vlp,Parnachev:2020zbr,Belin:2020lsr,Besken:2020snx,Karlsson:2021duj,Rodriguez-Gomez:2021pfh,Huang:2021hye,Rodriguez-Gomez:2021mkk,Krishna:2021fus,Korchemsky:2021htm,Bianchi:2021yqs,Karlsson:2021mgg,Huang:2022vcs,Dodelson:2022eiz,Dodelson:2022yvn, Bhatta:2022wga, Bajc:2022wws}.
In the context of AdS$_3$/CFT$_2$ such correlators have been well-studied in the literature, $e.g.$, \cite{Fitzpatrick:2014vua, Asplund:2014coa, Fitzpatrick:2015zha, Hijano:2015qja, Fitzpatrick:2015foa, Galliani:2016cai, Balasubramanian:2017fan, Galliani:2017jlg,  Faulkner:2017hll, Giusto:2018ovt, Giusto:2019pxc, Giusto:2020mup, Ceplak:2021wak, Bufalini:2022wyp, Berenstein:2022ico}.  
In $d=4$ holographic CFTs, heavy-heavy-light-light (HHLL) correlators were compared to thermal two-point functions in \cite{Kulaxizi:2018dxo} and 
several operator product expansion (OPE) coefficients of multi-stress tensor exchanges were computed in \cite{Fitzpatrick:2019zqz} which also observed that the scalar correlator in the near-lightcone limit is unaffected by higher-derivative interactions if one assumes a minimally coupled scalar in the bulk.  
Corrections to such universality due to non-minimally coupled interactions were discussed in \cite{Fitzpatrick:2020yjb}. 
In  \cite{Kulaxizi:2019tkd,Karlsson:2019dbd,Karlsson:2020ghx} the bootstrap procedure for computing HHLL correlators was developed.
Subsequently, it was pointed out  \cite{Huang:2021hye, Karlsson:2021mgg} that   higher-dimensional scalar correlators near the lightcone 
share certain similarities with the two-dimensional Virasoro vacuum blocks.  
Although the underlying mechanism responsible for this remains to be better understood, the time is ripe for investigation of a parallel story for the thermal correlators of stress tensors.  

An initial step towards this direction was made in \cite{Karlsson:2022osn}, which computed the thermal two-point correlators of stress tensors in $d=4$ holographic CFTs dual to Einstein gravity and read off conformal data beyond the leading order in the large $C_T$ expansion. 
 It was also observed that some OPE coefficients cannot be determined in the near-boundary analysis of the bulk equations of motion but these coefficients do not affect the near-lightcone $TT$ correlators.  
Subsequently, ref.  \cite{Huang:2022vet} included
the Gauss-Bonnet (GB) higher-derivative term in the gravitational action in $AdS_5$
to study what happens when the conformal collider bounds  \cite{Hofman:2008ar} in the dual CFT are saturated.\footnote{Conformal collider bounds, introduced in \cite{Hofman:2008ar}, are examples of   
Averaged Null Energy Conditions (ANECs) which  were shown to hold in unitary QFTs in \cite{Faulkner:2016mzt,Hartman:2016lgu}.}
It was shown that the thermal stress-tensor correlators near the lightcone 
take the vacuum form when this happens.

In this paper we point out that the stress-tensor correlators computed in Einstein-Gauss-Bonnet gravity  suggest a certain near-lightcone universality, which does {\it not} require ANEC saturation.  
We shall elaborate on this observation in more detail, but the main message is the following: thermal $TT$ correlators near the lightcone are completely determined by three universal functions (depending on polarization).
The Gauss-Bonnet term in the bulk Lagrangian only affects the arguments of these functions via corrections to the cubic stress-tensor couplings and the thermal stress-tensor one-point function.
We hypothesize that this remains the case in more general higher-derivative gravitational theories.\footnote{We expect similar results for the thermal correlators of conserved currents of spin one, with
two (instead of three) universal functions. }

The analysis in this paper involves  correlators in momentum space, and we mostly focus on retarded correlators in the near-lightcone regime.
To show that they have a universal structure, we identity a suitable  limit in momentum space
and show that the equations of motion of gravitational fluctuations in this limit take the same form as those in Einstein gravity.   
(These reduced bulk  equations of motion isolate the contributions of the leading-twist operators to the thermal $TT$ correlators in the dual CFTs.)
The near-lightcone $TT$ correlators depend on a single parameter, $\alpha \sim  q^+ (q^-)^3/T^4$, where $q^{\pm}$ are the lightcone momenta and $T$ is the temperature.
The expansion in the inverse powers of $\alpha$ is essentially an OPE
 (where only the leading-twist multi-stress tensor operators contribute because of the near-lightcone limit).
The reduced equations can be solved perturbatively in $1/\alpha$  and one can read off the corresponding OPE data.
Using  the WKB approximation, we  observe a non-perturbative imaginary term $\sim i e^{-\alpha^{1\over4}}$ in the retarded correlator.   
Such non-perturbative terms take the same  form irrespective of  polarization.

\subsubsection*{Outline}

The rest of this paper is organized as follows.  
In Section \ref{sec.unistat}, we   argue for the universality of thermal $TT$-correlators in the near-lightcone regime.
We then adopt a momentum-space approach in Section \ref{sec:Momentum}, where 
we show that the momentum-space equations of motion in Einstein-Gauss-Bonnet gravity in a suitable limit take the same form as the ones in Einstein gravity.  
By studying the on-shell action, we find that the near-lightcone $TT$ correlators (with three independent polarizations) in the holographic Einstein-Gauss-Bonnet theory are completely determined by three universal functions. 

In Section \ref{sec.lceq}, we compute  the near-lightcone correlators in momentum space perturbatively in $1/\alpha$.
In Section  \ref{sec.nonperts}, we  compute the near-lightcone correlators numerically and analyse the large $\alpha$ behavior using the WKB approach.
We extract the non-perturbative  term in the retarded correlators.
We discuss our results and pose some questions for future work  in Section \ref{sec.discussion}. 

In Appendix \ref{ap.pertpos}, we discuss the position-space correlators and perform the Fourier transform to check  the momentum-space results. 
We also estimate the radius of convergence in momentum space (for the scalar channel) of the near-lightcone correlators and find that the radius of convergence approaches zero, $i.e.$, the series is asymptotic.
Thermal conformal blocks in momentum space, discussed in Appendix  \ref{apB}, provide additional checks.  We list the equations of motion in Einstein-Gauss-Bonnet gravity in Appendix \ref{ap.eomsGB}. 


\section{Position-Space Correlators}\label{sec.unistat}

In this section, we point out a universality of the near-lightcone $TT$ correlators based on the position-space computation performed in \cite{Huang:2022vet}. 

Let us first recall that the vacuum stress-tensor three-point function in $d=4$ CFT in general depends on three coefficients $(\hat{a},\hat{b},\hat{c}$)  \cite{Osborn:1993cr}.  
 Depending on three different channels, $i.e.$, scalar, shear, and sound channels (polarisations),  the ``conformal collider bounds" place constraints 
 on some linear combinations of these coefficients \cite{Hofman:2008ar}: 
\begin{align}  \label{eq:CCB1}
		{\cal{C}}_{\rm scalar}&\equiv\frac{5\pi^2}{3C_T} \big(-7\aa-2\bb+\cc \big)\geq 0 \ , \\
  \label{eq:CCB2}
                  {\cal{C}}_{\rm shear}&\equiv \frac{10\pi^2}{3C_T}\big(16\aa+5\bb-4\cc \big) \geq 0\ , \\
  \label{eq:CCB3}
		{\cal{C}}_{\rm sound}&\equiv\frac{15\pi^2}{C_T} \big(-4\aa-2\bb+\cc\big)\geq0 \ ,
 \end{align}
where the central charge can be expressed as
\begin{align}
\label{CTabc}
C_T= \frac{\pi^2}{3}\big(14\aa-2\bb-5\cc\big) \ . 
\end{align} 

In this paper we are interested in thermal stress-tensor two-point functions on spatial $\mathbb{R}^3$ which involve three independent polarizations, mapping separately to the three different conformal collider bounds. 
The single-stress-tensor exchange contribution to thermal $TT$ correlators near the lightcone,  as shown by \cite{Kulaxizi:2010jt} using the stress-tensor OPE, are directly proportional to the same combinations of $(\hat{a},\hat{b},\hat{c})$ appearing in the conformal collider bounds.\footnote{One can use this fact and some minor assumptions to prove conformal collider bounds \cite{Komargodski:2016gci}.} At higher orders in the OPE, in large-$C_T$ CFTs, the thermal $TT$ correlators receive contributions from multi-stress-tensor exchanges.  
 In \cite{Karlsson:2022osn}, the OPE limit of the thermal $TT$ correlators in holographic CFTs dual to Einstein gravity was studied using the thermal conformal blocks, including double-stress tensors $\sim [T^2]_J$ with dimension $8+ \OO(1/C_T)$  and the corresponding CFT data was read off by comparison with a bulk computation.  
 
 It is interesting to ask what happens when one includes higher-derivative corrections to Einstein-gravity which modifies, in particular, the stress-tensor OPE coefficients $(\aa,\bb,\cc)$. In the context of holographic CFTs dual to Einstein-Gauss-Bonnet gravity,  the thermal $TT$ correlators were examined in \cite{Huang:2022vet} in position space.  
It was observed that, in the near-lightcone regime, the saturation of a CCB, $i.e.$, ANEC saturation, implies that the corresponding correlator takes the vacuum form, independent of temperature. 
To gain a broader understanding of the near-lightcone dynamics of the thermal correlators and their possible universal behavior, in this paper we study these thermal correlators away from  ANEC saturation.\footnote{In particular, the results of this paper are applicable for small values of the Gauss-Bonnet coupling, which one expects for a unitary theory with a large gap   \cite{Camanho:2014apa}.}  

Before proceeding, let us set up some notations. 
In any CFT, the thermal one-point function of an operator ${\cal{O}}_{\Delta,J}$ on $S^1_\beta\times\mathbb{R}^3$ with dimension $\Delta$ and spin $J$ is fixed, see $e.g.$ \cite{El-Showk:2011yvt,Iliesiu:2018fao}  
\begin{equation}
	\langle {\cal{O}}_{\Delta,J}\rangle_\beta = {b_{{\cal{O}}_{\Delta,J}}\over \beta^{\Delta}}(e_{\mu_1}\ldots e_{\mu_J}-\text{traces}) 
\end{equation}
where $\beta$ is the inverse temperature and $e^\mu$ is a unit vector on the thermal circle $S^1_\beta$. For our discussion of the near-lightcone correlators, it will be further useful to define the following quantities: 
\begin{align}
	\label{hatC}
	{\hat{\cal{C}} }_{\rm scalar}\equiv \frac{b_{T
	}}{C_T} \cal{C}_{\rm scalar}  \ , ~~~ 
	{\hat{\cal{C}} }_{\rm shear}\equiv \frac{b_{T
	}}{C_T}  \cal{C}_{\rm shear}\ , ~~~
	{\hat{\cal{C}} }_{\rm sound}\equiv \frac{b_{T
	}}{C_T} \cal{C}_{\rm sound} \ ,
\end{align} where  $\cal{C}_{\rm scalar, shear, sound}$ are defined in \eqref{eq:CCB1} -- \eqref{eq:CCB3}. 

We  will study the stress-tensor correlators integrated over the $xy$-plane:\footnote{See \cite{Karlsson:2022osn, Huang:2022vet} for related discussions on the integrated correlators.} 
\begin{equation}
	G_{\mu\nu, \rho\sigma}(t,z)\equiv \int_{\mathbb{R}^2}dxdy~ \langle T_{\mu\nu}(t,z, x,y)T_{\rho\sigma}(0)\rangle_\beta  \ . 
\end{equation}   The stress-tensor correlators 
can be classified into three independent channels, see, $e.g.$, \cite{Kovtun:2005ev}.  
 In the lightcone coordinates $x^{\pm}=t\pm z$ in Lorentzian signature, we consider the limit  $x^-\to 0$, with fixed $x^-(x^+)^3 \beta^{-4}$.
  In any four-dimensional CFT with no operators with twist less than or equal to two other than the stress-tensor, the thermal $TT$ correlators in the near-lightcone limit are given by 
{\small
	\begin{align} \label{eq:univ1}
	G_{xy,xy}(x^+,x^-)&=\frac{-\pi C_T}{10(x^-)^3(x^+)^3}\Big(1-\frac{{\hat{\cal{C}} }_{\rm scalar}}{\pi^2}\frac{x^-(x^+)^3}{\beta^4}+\ldots\Big) \ ,\\ 
\label{eq:univ2}
		G_{tx,tx}(x^+,x^-) &= \frac{-3\pi C_T}{80(x^-)^4(x^+)^2}\Big(1-\frac{{\hat{\cal{C}} }_{\rm shear}}{3\pi^2}\frac{x^-(x^+)^3}{\beta^4}+\ldots\Big)\ ,\\
\label{eq:univ3}
		G_{tz,tz}(x^+,x^-) &= \frac{-\pi C_T}{20(x^-)^5x^+}\Big(1-\frac{{\hat{\cal{C}} }_{\rm sound}}{12\pi^2}\frac{x^-(x^+)^3}{\beta^4}+\ldots\Big) \ .
\end{align}}The leading and subleading terms  in \eqref{eq:univ1} -- \eqref{eq:univ3} are universal, but the higher-order terms  {\it a priori} are model-dependent.  These higher-order terms contain contributions from the operators with larger dimensions -- in holographic CFTs, they include multi-stress tensor operators denoted as $[T^k]_J$.  

\subsubsection*{Position-Space Correlators in Holographic Einstein-Gauss-Bonnet Theory}

Here we make an observation based on the  thermal $TT$ correlators obtained holographically using Einstein-Gauss-Bonnet gravity \cite{Huang:2022vet}. 
Denote the dimensionless Gauss-Bonnet coupling as $\lambda_{\rm GB}$.\footnote{Einstein-Gauss-Bonnet gravity is discussed in more detail in Section \ref{sec:Momentum}.} We introduce a parameter 
\begin{equation}
\label{definek}
\kappa= \sqrt{1-4 \lambda_{\rm GB}}\ , 
\end{equation} which will help simplify expressions. The limit $\kappa \to 1$ recovers Einstein gravity. 
The coefficients $(\hat{a}, \hat{b}, \hat{c})$ can be related to the Gauss-Bonnet coupling: 
\begin{align} 
\label{abckappa}
\hat{a}=\frac{8C_T}{45\pi^2}(-6+\frac{5}{\kappa})\ ,~~~ \hat{b}=\frac{C_T}{90\pi^2}(33-\frac{50}{\kappa})\ , ~~~ \hat{c}=\frac{2C_T}{45\pi^2}(-84+\frac{61}{\kappa}) \ .
\end{align} 
The conformal collider bounds, \eqref{eq:CCB1} -- \eqref{eq:CCB3} translate to, respectively,  
\begin{align} 
\label{CCBkappa}
(5\kappa-4)\geq 0 \ ,~~~ (2-\kappa)\geq0 \ ,~~~  (4-3\kappa)\geq0 \ . 
\end{align}   

Now we make the following observation: the near-lightcone $TT$ correlators in holographic CFTs dual to Einstein-Gauss-Bonnet gravity computed in \cite{Huang:2022vet} can be recast in the following way:
{\small\begin{align}
	&G_{xy,xy}=\frac{-\pi C_T}{10(x^-)^3(x^+)^3}\left(1+\frac{1}{\pi^2}\bar{\alpha}_{\rm scalar}+\frac{5}{3\pi^4}\bar{\alpha}_{\rm scalar}^2+\ldots\right)\label{ob1} \ , \\        
	&G_{tx,tx}=\frac{-3\pi C_T}{80(x^-)^4(x^+)^2}\left(1+\frac{1}{3\pi^2}\bar{\alpha}_{\rm shear}+\frac{85}{504 \pi ^4}\bar{\alpha}_{\rm shear}^2+\ldots\right)\label{ob2} \ , \\
	&G_{tz,tz}=\frac{-\pi C_T}{20(x^-)^5x^+}\left(1+\frac{1}{12\pi^2}\bar{\alpha}_{\rm sound}+\frac{11}{756 \pi ^4}\bar{\alpha}_{\rm sound}^2+\ldots\right)\label{ob3} \ ,
\end{align}}where
\begin{align}\label{def:alphabar}
	\Big(  \bar{\alpha}_{{\rm scalar}},  \bar{\alpha}_{{\rm shear}}, \bar{\alpha}_{{\rm sound}} \Big)\equiv& \Big({\hat{\cal{C}} }_{\rm scalar},{\hat{\cal{C}} }_{\rm shear},{\hat{\cal{C}} }_{\rm sound} \Big )~ \frac{(-x^-(x^+)^3)}{\beta^4} \ .
\end{align}
While the first and second term are fixed by conformal symmetry as was shown in \eqref{eq:univ1} -- \eqref{eq:univ3}, one can see that, even at  $\cal O(\bar{\alpha}^2)$, the  dependence on the Gauss-Bonnet coupling can be  absorbed into the parameters $\bar{\alpha}$, which depend on the thermal one-point function of the stress tensor ($\sim b_T \beta^{-4}$) and $(\hat{a},\hat{b},\hat{c})$ through the particular combinations appearing in the conformal collider bounds.\footnote{The coefficients appearing in \eqref{ob1}-\eqref{ob3} in the $\bar{\alpha}$ expansion can also be verified to be the same in Einstein gravity \cite{Karlsson:2022osn} and Einstein-Gauss-Bonnet gravity \cite{Huang:2022vet}.} 
This observation hints at the following intriguing possibility: {\it multi-stress tensor contributions to the near-lightcone $TT$ correlators in holographic CFTs might be fixed by (the $k$-th power of) the single-stress-tensor contribution.}  
Exploring such a possibility and its consequences is the underlying motivation of the present work. 

To see this universality in position space we note that near the lightcone the corresponding reduced EoMs in Einstein-Gauss-Bonnet gravity obtained in Sec.\ 3.2 of \cite{Huang:2022vet} are identical to the ones in Einstein gravity, after performing suitable rescalings of the coordinates. 
This is easy to see in the scalar and shear channels, while the sound channel is technically more complicated when analyzed  in position space.  
In the next section we shall analyze the EoMs in  momentum space and show the universality in all channels.

Before moving to the momentum-space analysis, let us conclude this section with a technical remark: 
the structure of the higher-order terms, denoted by dots in \eqref{ob1} --  \eqref{ob3},  
in fact slightly differs from the first three terms we listed -- besides the corresponding dependence of $\bar{\alpha}_{\rm scalar}$, $\bar{\alpha}_{\rm shear}$ and $\bar{\alpha}_{\rm sound}$, the higher-order terms are multiplied by a $\log(-x^+x^-)$ piece. These arise for two different reasons, one is a contribution due to the anomalous dimensions of the multi-stress tensor operators and the other is because we consider the integrated correlator. We discuss this more in Appendix \ref{ap.pertpos} where we perform the Fourier transformation of the position-space correlators.


\section{Momentum-Space Correlators}\label{sec:Momentum}

In this section, we will study momentum-space  thermal $TT$ correlators and 
show that the near-lightcone correlators computed in the holographic Einstein-Gauss-Bonnet 
theory are universal: they are determined by three universal functions, depending on three polarizations.
We will hypothesize that this might be the case for all holographic theories. In subsequent sections, we will compute these functions.

Let us give a brief review on Einstein-Gauss-Bonnet gravity. The action in five dimensions is given by
{\small\begin{align}
S_{GB} = \frac{1}{16 \pi G} \int d^5 x \sqrt{-g} \left[{12\over L^2} + R  + \lambda_{\rm GB}   \frac{L^2}{2} 
\left( R^2 - 4 R_{\mu\nu} R^{\mu\nu} + R_{\mu\nu\rho\sigma} R^{\mu\nu\rho\sigma} \right) \right] \ .
\end{align}}The theory admits a black-hole solution \cite{Boulware:1985wk, Cai:2001dz}:
    \begin{equation}\label{gbbh}
        \dd s^2=\frac{r^2}{L^2}\left(-\frac{f(r)}{f_\infty}\dd t^2+\dd x^2+\dd y^2+\dd z^2\right)+\frac{L^2}{r^2}\frac{\dd r^2}{f(r)} 
    \end{equation}
where $f(r)$ and $f_\infty$ are 
    \begin{equation}\label{eq.fdefs}
        f(r)=\frac{1}{2\lambda_{GB}}\left[1-\sqrt{1-4\lambda_{GB}\left(1-\frac{r_+^4}{r^4}\right)}\right]\quad\text{and}\quad
        f_\infty=\frac{1-\sqrt{1-4\lambda_{GB}}}{2\lambda_{GB}} \ .
    \end{equation} 
The parameter $r_+$ is the location of the black-hole horizon. We focus on a planar horizon. In the following we set the AdS radius $L/\sqrt{f_\infty}$ to 1, $i.e.$ $L=\sqrt{f_\infty}$

To study the equations of motion of gravitational fluctuations, we consider the metric perturbation
$h_{\mu\nu} = h_{\mu\nu} (r) e^{-i\omega t + i q z}$, with the momentum along the $z$-direction, 
and adopt gauge-invariant quantities following the recipe in \cite{Kovtun:2005ev}. The radial gauge $h_{r\mu} = 0$ is used. 
The fluctuations are classified into three independent channels: scalar, shear, and sound.   
In Einstein-Gauss-Bonnet gravity, the corresponding gauge invariants are given by \cite{Buchel:2009sk}:
\begin{align}
  &Z^{\rm (GB)}_{\rm{scalar}} = \frac{1}{r^2}h_{xy}\ ,   \\
  &Z^{\rm (GB)}_{\rm{shear}} = \frac{\qf}{r^2}h_{tx} + \frac{\wf}{r^2}h_{zx} \ , \\
 &Z^{\rm (GB)}_{\rm {sound}} = \frac{2\qf^2}{r^2}h_{tt}+\frac{4\wf\qf}{r^2}h_{tz}+\frac{2\wf^2}{r^2}h_{zz}+\qf^2\left(\frac{2f+rf'}{2f_{\infty}}-\frac{\wf^2}{\qf^2}\right)\left(h_{xx}+h_{yy}\right) \ , 
\end{align}
where $(\wfr, \qfr) = {1\over {2\pi T}} (\omega, q)$ are the dimensionless frequency and momentum; $T$ is the Hawking temperature. 
The linearized equations of motion of gravitational fluctuations in Einstein-Gauss-Bonnet gravity were worked out in \cite{Buchel:2009sk}. 
(See \cite{Brigante:2007nu,Brigante:2008gz,Buchel:2009tt,deBoer:2009pn,Camanho:2009vw,Buchel:2010wf,Cai:2010cv,Bu:2015bwa,Grozdanov:2016vgg, Andrade:2016yzc, Grozdanov:2016zjj,Andrade:2016rln,Grozdanov:2016fkt,Casalderrey-Solana:2017zyh,Chen:2018nbh,An:2018dbz,Grozdanov:2021gzh}
for more recent applications of Einstein-Gauss-Bonnet holographic gravity.) 
In this work, we are  interested in the near-lightcone regime of the correlators.

Defining a coordinate $u = r_+^2/r^2$,  the equation of motion in each of the three channels can be written as a second-order differential equation:
\begin{align}
\label{ZEoM}
Z''(u) + A ~Z'(u) + B~ Z(u) = 0 \ .
\end{align}
The channel-dependent coefficients $A$ and $B$ are given in Appendix \ref{ap.eomsGB}.

\subsection{Equations of motion in the near-lightcone limit}

In this work, we are interested in the near-lightcone limit.  Denote $\mathfrak{q}^{\pm}= \wfr \pm \qfr$. We consider the following  limit:\footnote{We use the minus sign in the expression for $\alpha$ to make calculations in the space-like regime (which we will be interested in) more convenient.}
\be
\label{LCbulklimit}
   \mathfrak{q}^+ \ra \infty,  \qquad \mathfrak{q}^- \ra 0, \qquad \alpha = - \mathfrak{q}^+(\mathfrak{q}^-)^3 ~{\rm fixed},  \qquad \tilde u = u/ (\mathfrak{q}^-)^2  ~{\rm fixed}.
\ee
This limit isolates contributions from the leading twist operators, which corresponds to zooming in on the
near-boundary region in the bulk. For the position-space stress-tensor correlators, the corresponding near-lightcone limit in the bulk was discussed in \cite{Huang:2022vet}.

We next show that, in the limit \eqref{LCbulklimit}, the equations of motion in Einstein-Gauss-Bonnet gravity reduce to those in Einstein gravity. 
\\

\noindent {\bf Scalar Channel:} First we derive the reduced equation of motion in the scalar channel. 
In the limit \eqref{LCbulklimit}, the equation of motion at  leading order can be written as 
\begin{align}
\label{scalarint}
\alpha  (\kappa +1)^2 \Big(\alpha  \left(5 \kappa ^2+\kappa -4\right) \tilde u^2-8 \kappa ^2\Big)Z_{\rm scalar}(\tilde u)  -32 \kappa ^2 \left(Z'_{\rm scalar}(\tilde u)-\tilde u Z''_{\rm scalar}(\tilde u)\right)= 0 \ .
\end{align}
The observation is that, after rescaling the variables
\begin{align}
\label{rescalescalar}
\tilde u  = \frac{\kappa ^2 (\kappa +1)}{2 (5 \kappa -4)}  \tilde u_r   \ , ~~~
\alpha =  \frac{8 (5 \kappa -4)}{\kappa ^2 (\kappa +1)^3}  \alpha_r \ ,
\end{align} 
the equation of motion \eqref{scalarint} becomes 
\begin{align}
Z''_{\rm scalar}(\tilde u_r)  -\frac{1}{\tilde u_r} Z'_{\rm scalar}(\tilde u_r)+  \big(   \frac{ \alpha_r^2  \tilde u_r}{4 }  - { \alpha_r \over  \tilde u_r}  \big)Z_{\rm scalar}(\tilde u_r) = 0   \ .
\end{align}
Since this equation is completely independent of $\kappa$, we conclude it is identical to the equation of motion in Einstein gravity in the same limit.   
\\\\
\noindent {\bf Shear Channel:}  The equation of motion in the limit \eqref{LCbulklimit} is 
{\small
\begin{align}\label{uteq1}
&\alpha  (\kappa +1)^2  \left(8 \kappa ^2+\alpha  (\kappa -2) (\kappa +1) \tilde u^2\right)^2 \!Z_{\rm shear}(\tilde u) \\
&+ 32 \kappa ^2 \Big( \!\left(8 \kappa ^2+3 \alpha  (\kappa -2) (\kappa +1) \tilde u^2\right)\! Z'_{\rm shear}(\tilde u)\!+\! \left(\alpha  \left(\!-\kappa ^2+\kappa +2\right) \tilde u^2\!-8 \kappa ^2\right) \tilde u  Z''_{\rm shear}(\tilde u)\Big) \!= 0\,.  \nn
\end{align}}After performing the rescalings
\begin{align}
\label{rescaleshear}
\tilde u  = -\frac{ \kappa ^2 (\kappa +1)}{2 (\kappa -2)} \tilde u_r   \ , ~~~ \alpha =  -\frac{8 (\kappa -2)}{\kappa ^2 (\kappa +1)^3} \alpha_r \ , 
\end{align} 
we find 
{\small
\begin{align}
Z''_{\rm shear}(\tilde u_r)
- {1\over \tilde u_r }  \Big(\frac{3 \alpha_r \tilde u_r^2-4}{\alpha_r \tilde u_r^2-4 }  \Big) Z'_{\rm shear}(\tilde u_r)
+ \big(   \frac{ \alpha^2  \tilde u_r}{4 }  - { \alpha_r \over  \tilde u_r}  \big)  Z_{\rm shear}(\tilde u_r) = 0   
\end{align}}which is identical to the equation of motion in Einstein gravity in the same limit.
\\\\
\noindent{\bf Sound Channel:}  In the limit \eqref{LCbulklimit}, the sound-channel equation becomes  
{\small
\begin{align}\label{uteq2}
&\alpha  (\kappa +1)^2 \left(8 \kappa ^2+\alpha  (\kappa +1) (3 \kappa -4)  \tilde u^2\right) \left(24 \kappa ^2+\alpha  (\kappa +1) (3 \kappa -4) \tilde u^2\right) Z_{\rm sound}(\tilde u)\\
&+32 \kappa ^2 \Big(\left(24 \kappa ^2+5 \alpha  (\kappa +1) (3 \kappa -4)  \tilde u^2\right) Z'_{\rm sound}(\tilde u)-\left(\alpha  (\kappa +1) (3 \kappa -4) \tilde u^2+24 \kappa ^2\right) \tilde u  Z''_{\rm sound}(\tilde u)\Big) = 0  \ .\nn
\end{align}}Performing the rescalings
\begin{align}
\label{rescalesound}
\tilde u   = \frac{\kappa ^2 (\kappa +1)}{8-6 \kappa }  \tilde u_r   \ , ~~~
\alpha =  -\frac{8 (3 \kappa -4)}{\kappa ^2 (\kappa +1)^3} \alpha_r  \ , 
\end{align} 
we obtain the same equation as in Einstein gravity:
{\small
\begin{align}
Z''_{\rm sound}(\tilde u_r)
- {1\over \tilde u_r }  \Big(   \frac{5 \alpha_r  \tilde u_r^2-12}{\alpha_r  \tilde u_r^2-12} \Big) Z'_{\rm sound}(\tilde u_r)
+\big(   \frac{ \alpha_r^2  \tilde u_r}{4 }  - { \alpha_r \over  \tilde u_r}  \big)Z_{\rm sound}(\tilde u_r) = 0   \ .
\end{align}}In the Einstein gravity case, $\kappa=1$, one can verify that $\tilde u= \tilde u_r$, $\alpha= \alpha_r $ in all channels.

In summary, the momentum-space reduced equations of motion in the three different channels can be written as
\begin{align}
\label{rEoMmi}
&Z''(\tilde u_r) - {K(\alpha_r, \tilde u_r) \over \tilde u_r }  Z'(\tilde u_r) +\big(   \frac{ \alpha_r^2  \tilde u_r}{4 }  - { \alpha_r \over \tilde u_r}  \big) Z (\tilde u_r) = 0 
\end{align}
where  $K(\alpha_r, \tilde u_r)$ is channel-dependent:
\begin{align}
&~~~~~~~ \Big( K_{\rm scalar}, K_{\rm shear}, K_{\rm sound} \Big) = \Big(1,   \frac{3 \alpha_r \tilde u_r^2-4}{\alpha_r \tilde u_r^2-4 },    \frac{5 \alpha_r  \tilde u_r^2-12}{\alpha_r  \tilde u_r^2-12} \Big)  \ .
\end{align}
Hence, we have observed that the reduced equations of motion in the holographic Einstein-Gauss-Bonnet theory take the same form as the ones obtained in Einstein gravity.

We will next analyze the action and show that the near-lightcone $TT$ correlators are determined by three universal functions which correspond to the three independent polarizations.

\subsection{Thermal correlators from holography}

Here we compute holographic thermal $TT$ correlators $G_{\mu\nu,\rho\lambda}$  in four spacetime dimensions.  
The symmetries of the theory imply that the momentum-space retarded correlator 
has the following form \cite{Kovtun:2005ev}:
    \begin{equation}
    G_{\mu\nu,\rho\lambda}(k)=L_{\mu\nu,\rho\lambda}G_{\rm scalar}(k)+ S_{\mu\nu,\rho\lambda}G_{\rm shear}(k)+Q_{\mu\nu,\rho\lambda}G_{\rm sound}(k) \ ,
    \end{equation}
where $G_{\rm scalar}$, $G_{\rm shear}$ and $G_{\rm sound}$ are three independent scalar functions of momenta and the tensor structures $L_{\mu\nu,\rho\lambda}$, $S_{\mu\nu,\rho\lambda}$ and $Q_{\mu\nu,\rho\lambda}$ are fixed by the symmetries, see, $e.g.$, \cite{Kovtun:2005ev}.
If not stated otherwise, we use  Minkowski signature.

Following the Lorentzian AdS/CFT dictionary \cite{Policastro:2001yc,Son:2002sd,Policastro:2002se,Policastro:2002tn,Kovtun:2004de}
we impose incoming boundary conditions  at the horizon, $\tilde u \ra \infty$.
To compute the correlators we need the $\OO(Z^2)$ on-shell action in all three channels, which  is given by \cite{Buchel:2009sk} 
{\small  
    \begin{align}
        I_{\rm scalar}&=-\frac{\pi^2C_T r_+^4}{80 f_\infty^2}\lim_{u\rightarrow0}\int\frac{\dd\omega\dd q}{(2\pi)^2}\frac{1}{u}\pdv{u}\Zt(u,k)\Zt(u,-k) \ , \\
        I_{\rm shear}&=-\frac{\pi^2C_T r_+^4}{80 f_\infty^2}\lim_{u\rightarrow0}\int\frac{\dd\omega\dd q}{(2\pi)^2}\frac{1}{u(\wf^2-\qf^2)}\pdv{u}\Zv(u,k)\Zv(u,-k) \ , \\
        I_{\rm sound}&=\frac{3\pi^2C_T r_+^4}{320 f_\infty^2}\lim_{u\rightarrow0}\int\frac{\dd\omega\dd q}{(2\pi)^2}\frac{1}{u(3\wf^2-\qf^2(3-u^2))^2}\pdv{u}\Zs(u,k)\Zs(u,-k) \ ,
    \end{align}}
where $C_T$ is the central charge of the holographic Einstein-Gauss-Bonnet theory.

The correlators are given by  
\begin{equation}\label{eq.origG}
    G_{\rm scalar}=\frac{\pi^2C_T r_+^4}{10 f_\infty^2}\frac{\mathcal{B}_{\rm scalar}}{\mathcal{A}_{\rm scalar}} \ ,\,
    G_{\rm shear}=-\frac{\pi^2C_T r_+^4}{10 f_\infty^2}\frac{\mathcal{B}_{\rm shear}}{\mathcal{A}_{\rm shear}} \ ,\,
    G_{\rm sound}=-\frac{\pi^2C_T  r_+^4}{10 f_\infty^2}\frac{\mathcal{B}_{\rm sound}}{\mathcal{A}_{\rm sound}} 
    \end{equation}
where $\mathcal{A}$ and $\mathcal{B}$ are the coefficients in the near-boundary expansion:\footnote{This is the standard Frobenius expansion around $u=0$, $i.e.$, $Z=\mathcal{A}u^{\Delta_1}(1+\ldots)+\mathcal{B}u^{\Delta_2}(1+\ldots)$, where $\Delta_1$ and $\Delta_2$ are the leading exponents. For equations \eqref{scalarint}, \eqref{uteq1} and \eqref{uteq2} one finds $\Delta_1=0$ and $\Delta_2=2$. Since these differ by an integer, one has to include log-terms in the $\Delta_2$-part of the solution. Inserting the resulting ansatz into the corresponding equation, one can determine the subleading coefficients in terms of $\mathcal{A}$ and $\mathcal{B}$. } 
    \begin{equation}\label{eq.ten.schemufrom}
        Z(u)=\mathcal{A}-\frac{\alpha\mathcal{A}}{f^2_\infty(\mathfrak{q}^-)^2}u+\mathcal{B}u^2-\frac{\alpha^2\mathcal{A}}{2f^4_\infty(\mathfrak{q}^-)^4}u^2\log u+\ldots
    \end{equation}
Let us adopt a new variable  
    \begin{equation}
        x \equiv \tilde{u}_r \alpha_r   = f_\infty^{-2} \tilde u \alpha=-f_\infty^{-2}\mathfrak{q}^+\mathfrak{q}^- u\ ,
    \end{equation}
 where we used (\ref{rescalescalar}), (\ref{rescaleshear}), (\ref{rescalesound}) and
 the relation 
 \be
 \label{kappafinfty}
 \kappa = \frac{2}{f_\infty} -1 \ .
 \ee
We can now rewrite the equations of motion \eqref{rEoMmi} as 
\begin{align}
\label{eq.reommnew}
&Z''(x)- B (x, \alpha_r) Z'(x)+\frac{x^2-4\alpha_r}{4\alpha_r x}Z(x)=0 \ ,\\
\nonumber &~~~~~~~~~~\big(B_{\rm scalar}, B_{\rm shear} , B_{\rm sound} \big)= \Big( \frac{1}{x},   \frac{3x^2-4\alpha_r}{x^3-4\alpha_r x},\frac{5x^2-12\alpha_r}{x^3-12\alpha_r x}  \Big) \ .
\end{align}
Note that $x \to 0$ is the boundary limit while $x \to \infty$ corresponds to  the black hole horizon.

Before solving the equation \eqref{eq.reommnew}, let us consider a formal analysis of the near-boundary structure of the correlators. 
The near-boundary expansion up to quadratic order is the same in all channels and reads
    \begin{equation}\label{eq.ten.schemxfrom}
    Z(x)=a-ax+bx^2-\frac{a}{2}x^2\log x+\ldots 
    \end{equation} 
where $a$ and $b$ are functions of $\alpha_r$.  
The coefficients $\mathcal{A}$ and $\mathcal{B}$ can be related to $a, b$ in the following way:
    \begin{equation}\label{abexpansion}
        \mathcal{A}=a \ , ~~~  \mathcal{B}=(\mathfrak{q}^+\mathfrak{q}^-)^2 f_\infty^{-4} \left(b-\frac{a}{2}\log(-f_\infty^{-2}\mathfrak{q}^+\mathfrak{q}^-)\right).
    \end{equation}
We can now write the correlator in e.g.\ the scalar channel as
    \begin{equation}\label{eq.G3defab}
        G_{\rm scalar}=\frac{\pi^2C_T}{160} (q^+ q^-)^2  \left( \mathfrak{f}_{\rm scalar}(\alpha_r)-\frac{1}{2}\log(-q^+q^-)\right) \ ,
~~~     \mathfrak{f}_{\rm scalar}(\alpha_r)=\frac{b}{a} \ ,
    \end{equation}
where we have ignored terms analytic in momenta.\footnote{These correspond to contact terms in position space.} Note that the ratio $\mathfrak{f}=\frac{b}{a}$ only depends on 
$q^{\pm}$ through $\alpha_r$.  
The function $\ff_{\rm scalar}$ can be obtained via
    \begin{equation}\label{eq.fscalar}
        \mathfrak{f}_{\rm scalar}\left(\alpha_r\right) =\lim_{x\rightarrow0}\pdv[2]{x}\left(\frac{\Zt(x)}{2\lim_{x'\rightarrow0}\Zt(x')}+\frac{x^2}{4}\log x\right) 
    \end{equation} after we solve for $Z(x)$ from the corresponding equation of motion.

Analogously, in the other channels we obtain
    \begin{align}
        G_{\rm shear}&=-\frac{\pi^2C_T}{160} (q^+q^-)^2  \left( \mathfrak{f}_{\rm shear}(\alpha_r)-\frac{1}{2}\log(-q^+q^-)\right)\label{eq.G1defab}\ , \\
        G_{\rm sound}&=-\frac{\pi^2C_T}{160} (q^+q^-)^2  \left( \mathfrak{f}_{\rm sound}(\alpha_r)-\frac{1}{2}\log(-q^+q^-)\right)\label{eq.G2defab} \ .
    \end{align}
The functions $\ff_{\rm shear}(\alpha_r)$ and $\ff_{\rm sound}(\alpha_r)$ are again defined as the ratios of the corresponding coefficients in the near-boundary expansion of  $Z(x)$. 

In the next two sections, we will compute the functions $\ff_{{\rm scalar}, {\rm shear}, {\rm sound}}(\alpha_r)$  first perturbatively in $1/\alpha_r$ and then numerically. A few comments are in order:  

\begin{itemize}

\item We again emphasize that  the near-lightcone $TT$ correlators in the holographic Einstein-Gauss-Bonnet theory are expressed 
in terms of the same functions (there are three of them, which corresponds to the three independent polarizations) as the ones that 
appear in pure Einstein gravity.  It is possible that this universality holds true more generally, going beyond the holographic Einstein-Gauss-Bonnet theory. We rephrase it in the language of the 
stress-tensor three-point couplings below.

\item The function $\ff$ has a perturbative expansion in $1/\alpha$  
which is basically an OPE, and also non-perturbative terms of the type $e^{-\alpha^{1\over4}}$. 
The non-perturbative terms correspond to  tunneling under the potential barrier in the Schr\"odinger equation  
which can be obtained from (\ref{eq.reommnew})
and are sensitive to the boundary conditions at the horizon ($x{\ra} \infty$), as we explain in Section \ref{sec.nonperts}.

\item The perturbative expansion is not sensitive to the horizon boundary conditions -- it is equivalent to the OPE which can also be performed
in  position space.
In Appendix \ref{ap.pertpos}, we explicitly match several terms between the position- and momentum-space expansions.

\end{itemize}

Let us now point out that the first term in the perturbative expansion of  $\ff$ is a non-physical (cutoff-dependent) number, while
the second term (proportional to $\beta^{-4}$) is fixed by the $TTT$ three-point couplings and the coefficient $b_T$  \cite{Kulaxizi:2010jt}
\be
\label{reminderalpha}
\ff_{\rm scalar} (\alpha_r) |_{\beta^{-4} } \sim \frac{7 \aa +2 \bb -\cc}{14 \aa- 2\bb -5 \cc} \ \frac {{\tilde b}_T }{\alpha}
\ee  
where ${\tilde b}_T $ is defined by 
\be
\label{defbt}
\tilde b_T = \frac{b_T}{C_T} \ .
\ee
The ratio of  (\ref{reminderalpha}) to the corresponding term in Einstein gravity is given by
\be
\label{fratio}
\frac{   \ff_{\rm scalar} (\alpha_r) |_{\beta^{-4}}  }{   \ff_{\rm scalar} (\alpha) |_{\beta^{-4} } }=-5  \frac{7 \aa +2 \bb -\cc}{14 \aa- 2\bb -5 \cc}  \ \frac{\tilde b_T}{\tilde b_{T,0}}= \frac{8 (5 \kappa-4)}{ \kappa^2 (\kappa+1)^3}
\ee
where the zero in the subscript indicates the corresponding value for Einstein gravity.
The first equality in  (\ref{fratio}) follows from (\ref{reminderalpha}) and to get the second equality
we have used expressions for $\aa,\bb,\cc$ from (\ref{abckappa}). We also used\footnote{This relation can be obtained from, $e.g.$, Eq. (2.9) of \cite{Huang:2022vet}.  
	In this work we do not need to know the value of $\tilde b_{T,0}$, but it is available -- for example it can be read off from Eqs. (3.15-3.16) in \cite{Karlsson:2022osn}.}  
\be
\label{btgb}
\frac{\tilde b_T }{\tilde b_{T,0}}= \frac{8}{\kappa (\kappa+1)^3 } \ .
\ee 
Note that the first equality in (\ref{fratio}) was derived from the Ward identity.
Eq. (\ref{fratio})  is consistent with the relation between $\alpha$ and $\alpha_r$ in (\ref{rescalescalar}), as it should.
It is tempting to propose that in any holographic theory the function that enters the correlator in the
scalar channel is
\be
\label{genscalar}
\ff_{\rm scalar}  (\alpha_r) =  \ff_{\rm scalar}\left(\left[ -5  \ \frac{7 \aa +2 \bb -\cc}{14 \aa- 2\bb -5 \cc}  \ \frac{\tilde b_T}{\tilde b_{T,0}}  \right]^{-1} \alpha \right) \ .
\ee
This statement means that the near-lightcone correlator for all holographic CFTs  
is completely fixed in terms of basic conformal data (such as $\aa,\bb,\cc, \tilde b_T$) and the function $\ff$, which we compute in the next two sections. 
For the other channels, similar logic implies 
\be
\label{genshear}
\ff_{\rm {shear}}  (\alpha_r) =  
\ff_{\rm {shear}} \left( \left[   10\ \frac{ 16 \aa +5 \bb -4 \cc  }{ 14 \aa- 2\bb -5 \cc }   \ \frac{\tilde b_T}{\tilde b_{T,0}} \right]^{-1}  \alpha \right) \ ,
\ee
\be
\label{gensound}
\ff_{{\rm sound}}(\alpha_r) = 
\ff_{\rm {sound}}\left( \left[  -45\ \frac{  4 \aa +2 \bb -\cc  }{ 14 \aa- 2\bb -5 \cc  }  \ \frac{\tilde b_T}{\tilde b_{T,0}} \right]^{-1}   \alpha \right) \ .
\ee
Note that the combinations of  $\aa,\bb,\cc$ in the numerators are proportional to the corresponding ANECs.
Hence, one obtains the vacuum result once an ANEC gets saturated, reproducing the results of  \cite{Huang:2022vet}.

\section{Perturbative Analysis}\label{sec.lceq}

Here  we shall focus on computing the 
near-lightcone thermal $TT$ correlators assuming Einstein gravity in the bulk, setting $ \alpha_r = \alpha$.
We focus on the
scalar channel where the perturbative expansion reads
    \begin{equation}
        \ff_{\rm scalar}(\alpha) =\sum_{n=0}^\infty ~{\ff^{(n)}_{\rm scalar} \over \alpha^n} .
    \end{equation}
The computation in the other two channels is analogous, so we will simply list the corresponding results.  

    \subsection{Leading order (vacuum correlators) }
    
We start with the $\OO(1/\alpha^0)$ term, $\ff_{\rm scalar}^{(0)}$, in the large $\alpha$ expansion, which corresponds to the vacuum solution.  
As $\alpha\to\infty$ the reduced equation of motion becomes
    \begin{equation}\label{eq.0reomm3}
        \left(\pdv[2]{x}-\frac1x\pdv{x}-\frac1x \right)\Zt^{(0)}=0 \ ,
    \end{equation}
and admits an analytic solution in terms of the Bessel functions:
    \begin{equation}\label{homog}
\Zt^{(0)}(x)=2c_1xI_2\left(2\sqrt{x}\right)+2c_2xK_2\left(2\sqrt{x}\right) 
    \end{equation}  with  two coefficients $c_1$ and $c_2$.  
Regularity in the bulk requires $c_1=0$, while $c_2$ remains arbitrary. 
This remaining coefficient corresponds to the  norm of $\Zt^{(0)}$ which does not affect the value of the correlator; without loss of generality we require $a=1$ in \eqref{abexpansion}, thus $c_2=1$. 
The near-boundary expansion is then given by 
    \begin{equation}\label{kekeska}
       \Zt^{(0)}(x)=1-x+\frac{1}{4}x^2(3-4\gamma-2\log(x))+\frac{1}{36}x^3(17-12\gamma-6 \log (x))+\mathcal{O}\left(x^4\right) \ , 
    \end{equation}
where $\gamma$ is Euler's constant. From \eqref{kekeska} we find the ratio $\ff^{(0)}_{\rm scalar}= {b\over a} = \frac{1}{4}\left(3-4\gamma\right)$.
Thus, the $\OO(1/\alpha^0)$ contribution to the correlator $G_{\rm scalar}$ is
    \begin{equation}\label{eq.0corrG3}
       G^{(0)}_{\rm scalar}=\frac{\pi^2C_T}{160} (q^+q^-)^2\left(\frac{1}{4}\left(3-4\gamma\right)-\frac{1}{2}\log(-q^+q^-)\right) \ .
    \end{equation}
A similar calculation in the shear and sound channels yields the same result: 
    \begin{equation}\label{eq.analytic123}
\ff^{(0)}_{\rm scalar}=\ff^{(0)}_{\rm shear}=\ff^{(0)}_{\rm sound}=\frac{1}{4}\left(3-4\gamma\right).
    \end{equation}
Note that in all channels $\ff^{(0)}$ corresponds to the contact term.

    \subsection{Subleading order}\label{ssec.firstthorderm}

To proceed with perturbative expansion for the scalar channel, it will be useful to convert the corresponding reduced equation of motion \eqref{eq.reommnew} into  the following Schrödinger form:
    \begin{equation}\label{shok}
        \pdv[2]{x}Y(x) + \left(-\frac{3}{4x^2} - \frac{1}{x} + \frac{x}{4 \alpha}\right) Y(x) = 0 \ , ~~ Y(x)=\sqrt{\frac{\alpha}{x}}Z(x) \ .
    \end{equation}
The expansion in $1/\alpha$ reads 
    \begin{equation}\label{yeq}
        Y(x) = Y^{(0)}(x) + \frac{1}{\alpha} Y^{(1)}(x) + \dots
    \end{equation}
where  $Y^{(0)}(x)=\sqrt{\frac{\alpha}{x}}\Zt^{(0)}(x)$ was computed before. 
Expanding the equation \eqref{shok} to  $\OO(1/\alpha)$ gives 
\begin{equation}
 \pdv[2]{x}Y^{(1)}(x)-\left(\frac{1}{x} + \frac{3}{4 x^2} \right) Y^{(1)}(x) = - \frac{x}{4} Y^{(0)}(x) \ .
\end{equation}
The solution can be written in terms of the MeijerG functions:  
\begin{equation}
\begin{split}
Y^{(1)}(x)=
&\frac{x^3}{4 \sqrt{\pi }} \Bigg[\!K_2\left(2 \sqrt{x}\right) G_{1,3}^{2,1}\Bigg(4x\Bigg|
\begin{array}{c}
 1 \\
 \frac{1}{2},\frac{5}{2},-\frac{5}{2} \\
\end{array}
\Bigg)\!-\!\pi  I_2\left(2 \sqrt{x}\right) G_{2,4}^{3,1}\Bigg(4x\Bigg|
\begin{array}{c}
 -\frac{3}{2},1 \\
 -\frac{3}{2},\frac{1}{2},\frac{5}{2},-\frac{5}{2} \\
\end{array}
\Bigg)\Bigg]\\
&-2 i c_3 \sqrt{x} I_2\left(2 \sqrt{x}\right)+2 c_4 \sqrt{x} K_2\left(2 \sqrt{x}\right) \ . 
\end{split}
\end{equation}
Expanding the solution near the horizon, regularity restricts the coefficient $c_3 = \frac{i}{10}$. Setting $a=1$ in the expansion \eqref{abexpansion} leads  to $c_4 = 0$, which completely fixes the $\OO(1/\alpha)$ solution.
This gives the following contribution to the correlator
\begin{equation}
    G_{\rm scalar}=  G^{(0)}_{\rm scalar}+\frac{\pi^2 C_T}{1600} \ {(q^+q^-)^2 \over \alpha} + \OO(\alpha^{-2}).
\end{equation}
Following the same path, we also obtain the results in the shear and sound channels. 

Extracting the functions $\ff_{\rm scalar}$, $\ff_{\rm shear}$ and $\ff_{\rm sound}$, 
we find 
    \begin{align}\label{eq:TExch}
        \ff^{(1)}_{\rm scalar}=\frac{1}{10} \ , ~~~~
        \ff^{(1)}_{\rm shear}=-\frac{1}{40} \ , ~~~~
        \ff^{(1)}_{\rm sound}=\frac{1}{60}\ .
    \end{align} 

\subsection{Higher orders}

The general equation satisfied by the higher-order terms is given by
\begin{equation}
 \pdv[2]{x}Y^{(n)}(x)-\left(\frac{1}{x} + \frac{3}{4 x^2} \right) Y^{(n)}(x) = - \frac{x}{4} Y^{(n-1)}(x)\ .
\end{equation}
The homogeneous solution is given by
    \begin{equation}
Y_H^{(n)}(x) = -2 i c_{2n+1} \sqrt{x} I_2\left(2 \sqrt{x}\right)+2 c_{2n+2} \sqrt{x} K_2\left(2 \sqrt{x}\right),
    \end{equation}
while a particular solution can be expressed via the Green's function method
described in, $e.g.$, \cite{Romatschke_2009}, as
\begin{equation}
\begin{split}\label{ypn}
Y_P^{(n)}(x) = & (-2 i \sqrt{x} I_2(2\sqrt{x})) \int_0^x \dd y \frac{i}{2} (2 \sqrt{y} K_2(2\sqrt{y})) \frac{-y}{4} Y^{(n-1)}(y)\\
& + (2 \sqrt{x} K_2(2\sqrt{x})) \int_x^\infty \dd y \frac{i}{2} (-2 i \sqrt{y} I_2(2\sqrt{y})) \frac{-y}{4} Y^{(n-1)}(y) \ .
\end{split}
\end{equation}
Although it is not easy to perform the integrals in \eqref{ypn} explicitly, one can examine the near-horizon behaviour of the particular solution
which will be used to impose  regularity at the horizon
    \begin{equation}\label{obecnaformula}
Y_P^{(n)}(x \to \infty) = \left(\lim_{x\to \infty}-2 i \sqrt{x} I_2(2\sqrt{x})\right) \int_0^\infty \dd y \frac{i}{2} (2 \sqrt{y} K_2(2\sqrt{y})) \frac{-y}{4} Y^{(n-1)}(y)      \ .    
    \end{equation}

Focus now on the  $\OO(1/\alpha^2)$ term. Regularity of $Y^{(2)}(x)$ near the horizon leads to $c_5=\frac{i}{20}$, while requiring $a=1$ in the expansion \eqref{abexpansion} fixes the remaining coefficient $c_6 = 0$. 
This yields the $\OO(1/\alpha^2)$ contribution
\begin{equation}
 G_{\rm scalar}=  G^{(0)}_{\rm scalar}+\frac{\pi^2 C_T}{1600} \ {(q^+q^-)^2 \over \alpha} + \frac{\pi^2 C_T}{3200} \ {(q^+q^-)^2\over \alpha^2}+\OO(\alpha^{-3}).    
\end{equation}

In the same way, we can obtain the results in the remaining two channels. After extracting the function $\ff$, we find the following results:   
    \begin{align}
        \ff^{(2)}_{\rm scalar}=\frac{1}{20} \ , ~~~~
        \ff^{(2)}_{\rm shear}=-\frac{1}{40}\ , ~~~
        \ff^{(2)}_{\rm sound}=\frac{11}{420} \ . 
    \end{align} 
    
The same method in principle allows one to work out higher-order terms. 
In  Appendix \ref{ap.pertpos} we also verify the above results using the position-space approach.

{\it Radius of convergence:}  
\label{ssec.radiuscon}
Let us now estimate the radius of convergence of the perturbative expansion. We focus on the scalar channel. Define
\begin{equation}\label{rok}
	r_n=\abs{\frac{\ff^{(n)}_{\rm scalar}}{\ff^{(n+1)}_{\rm scalar}}} \ .
\end{equation}
We plot $r_n(n)$ in Fig.\ \ref{fig.rc} in Appendix \ref{ap.pertpos}.\footnote{The higher-order perturbative terms in this plot are computed in position space and then Fourier transformed to momentum space.}
The radius of convergence is defined as $\lim_{n\to\infty}  r_n$ and we find that it seems to be zero.


\section{Non-Perturbative Behavior} \label{sec.nonperts} 

In this section we analyze the stress-tensor correlators in all channels by solving the reduced equations of motion \eqref{eq.reommnew} numerically.
Note that the retarded correlators in general have an  imaginary part, which represents a purely non-perturbative contribution.
Using a WKB approximation we analyse this contribution explicitly and show that all three channels decay exponentially at the same rate. 

    \subsection{Numerical solution}

\begin{figure}[t!]
\begin{center}
\hspace*{-1.5cm}
\includegraphics[width=0.85\textwidth]{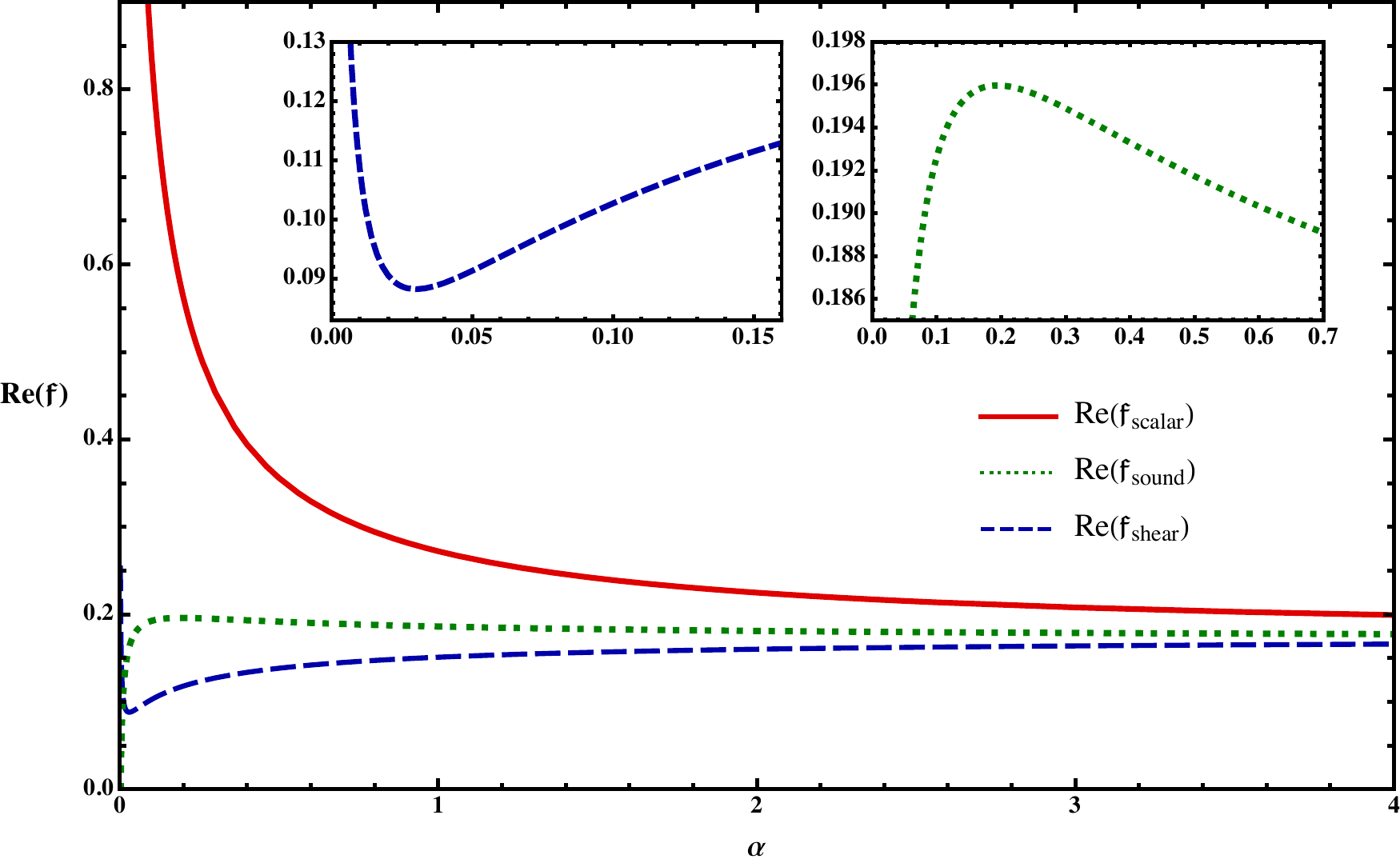}
\caption{The real part of $\mathfrak{f}(\alpha)$. 
The upper (solid, red) line corresponds to the scalar channel. 
The middle (dotted, green) line corresponds to the sound channel. 
The bottom (dashed, blue) line corresponds to the shear channel. 
 At large $\alpha$, all three lines approach the expected value, $\frac{3}{4}-\gamma$ $\approx 0.173$, where $\gamma$ is Euler's constant. 
The additional  smaller figures show the local minimum and maximum that appear in the shear and sound channels, respectively, in a small $\alpha$ region.}
\label{fig.re}
\end{center}
\end{figure}

\begin{figure}[ht]
\begin{center}
\hspace*{-1.5cm}
	\includegraphics[scale=0.65]{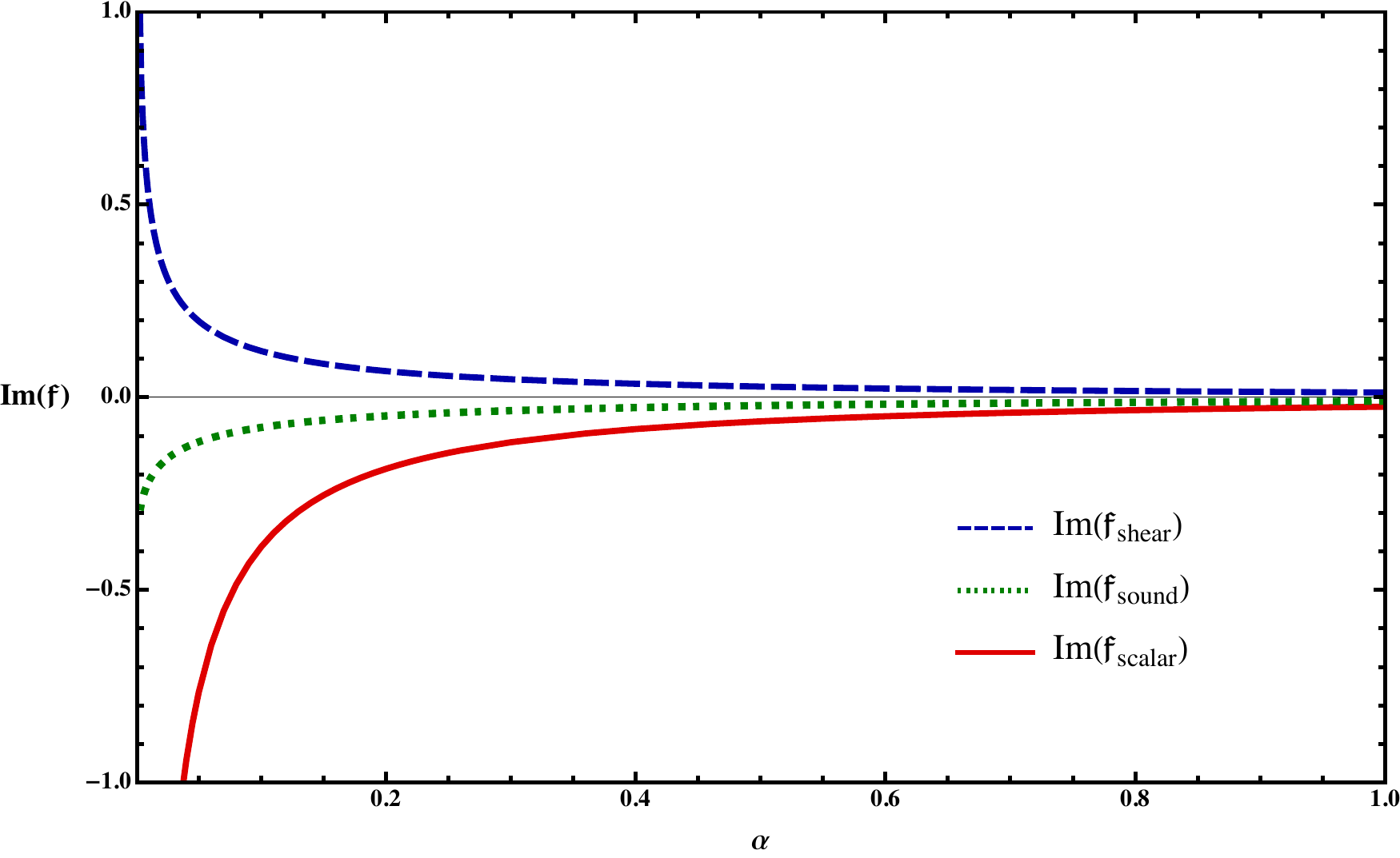}
    \caption{The imaginary part of $\mathfrak{f}(\alpha)$. 
The upper (dashed, blue) line corresponds to the shear channel. 
The middle (dotted, green) line corresponds to the sound channel. 
The bottom (solid, red) line corresponds to the scalar channel. }
	\label{fig.imAll}
\end{center}
\end{figure} 

In what follows, we again focus on space-like momenta where $\alpha$ (and thus $x$) is positive. In general, the solution in the limit $x\rightarrow\infty$ is a regular function multiplied by a superposition of incoming and outgoing waves. 
The natural choice is to pick the incoming-wave condition as discussed in \cite{Policastro:2001yc,Policastro:2002se,Policastro:2002tn,Son:2002sd}. 
With this choice one obtains the retarded correlators in the dual CFT. 

Consider the scalar channel.  Near the horizon the corresponding reduced equation \eqref{eq.reommnew} reduces to 
    \begin{equation}\label{eq.nearh.reomm3}  
       \left( \pdv[2]{x}-\frac1x\pdv{x}+\frac{x}{4\alpha} \right)\Zt(x\!\rightarrow\!\infty)=0 
    \end{equation}
where $\Zt(x\!\rightarrow\!\infty)$ denotes the solution deep in the bulk.
This equation can be solved analytically in terms of the (differentiated) Airy functions. Expanding the solution for large $x$ and picking the incoming wave, one finds
    \begin{equation}\label{eq.rawsol3hor}
        \Zt(x\!\rightarrow\!\infty)=x^{\frac14}e^{-\frac{i x^{3\over 2}}{3 \sqrt{\alpha}}} \left(\alpha^{-{1\over12}}-\frac{7 i \alpha^{5\over 12}}{24x^{3\over 2}}+\ldots\right)\ .
    \end{equation}
We use this expression to  numerically solve equation \eqref{eq.reommnew} starting from  large values of $x$ all the way to the boundary at $x=0$.
Then, using  \eqref{eq.fscalar} we compute the function $\ff_{\rm scalar}$ from this numerical solution.
We compute $\ff_{\rm shear}$ and $\ff_{\rm sound}$ in a similar way.

We next present the numerical results of both the real and imaginary parts of the correlators, for all three channels.

{\bf{Real part:}} We plot the real part of the function $\ff(\alpha)$ in Fig.\ref{fig.re}. 
For large $\alpha$, in all channels the numerical solutions quickly converge to the value $\frac{3}{4}-\gamma$, $i.e.$, the leading order in the perturbative expansion. We have also verified that, for every $n\in\mathbb{N}$, there exists a value $\alpha_n$, such that for all $\alpha>\alpha_n$ the $n$-th order perturbative expansion approximates the numerical solution better than any expansion with $n-1$ (or less) terms.  

{\bf{Imaginary part:}} We present the numerical solutions of the imaginary part of the function $\ff(\alpha)$ in Fig.\ \ref{fig.imAll}.
The imaginary part of the correlators is {\it purely} non-perturbative in the $1/\alpha$ expansion. Its concrete form is sensitive to the boundary condition at the horizon.  
If one instead imposes the outgoing-wave condition in the bulk and computes an advanced correlator, one finds that the correlator has the same real part but the imaginary part differs by a sign
(this also  follows from the general properties of Green's functions).
\subsection{Imaginary part of  correlators from WKB }  \label{sec.wkbim} 

An interesting question is how to estimate the non-perturbative behavior of the thermal correlators calculated numerically above. Let us answer this question by calculating the decay rate of $\Im G_R$ using the WKB analysis. To do so, 
we shall transform the reduced equations of motion \eqref{eq.reommnew} to the following form:
    \begin{equation}
    \label{forQM}
    \hbar^2\dv[2]{Z(\xi)}{\xi}=V(\xi)Z(\xi)\ .
    \end{equation} 
Compared to the previous Schr{\"o}dinger-like equation \eqref{shok}, here the $V(\xi)$ term is independent of the expansion parameter, 
allowing us to perform a standard WKB analysis. 
Starting from \eqref{eq.reommnew},
we first rescale $x\rightarrow\sqrt{\alpha}y$ to have $B(x)=B(y)\alpha^{-1/2}$ where $B(y)$ is independent of $\alpha$. 
We next introduce $\xi(y)$ which satisfies the relation
    \begin{equation}\label{eqrk}
        \partial_y\log\big(\partial_y\xi\big)=B(y)\ .
    \end{equation}
The equations of motion can be written as
    \begin{equation}\label{fulrk}
        \alpha^{-\frac{1}{2}}Z''(\xi)=-\frac{y^2-4}{4y(\xi'(y))^2}Z(\xi)\ .
    \end{equation}
The parameter $\alpha$ plays the role of $\hbar$; the precise identification is $\alpha^{-{1\over 4}}=\hbar$.  
For simplicity we omit the channel index in $\xi$ and $Z$. 
Note that the potential  in \eqref{fulrk} is positive in the  region $y\in(0,2)$
which corresponds to a classically forbidden region. 

It may be useful to list  explicit expressions relating $y$ and $\xi$ for different channels:\footnote{Note \eqref{eqrk} is a second-order differential equation. We have chosen two integration constants such that these  expressions look simple.} 
    {\small\begin{align}
    \text{Scalar}&:\quad y=\sqrt{\xi} \ , \label{pozor0}\\
    \text{Shear}&:\quad y=\sqrt{4\pm\abs{\xi}^\frac12}\label{pozor} \ , \\
    \text{Sound}&:\quad y=\sqrt{12+\xi^{1\over 3}}\ . \label{pozor2}
    \end{align}}In \eqref{pozor} we use plus for $\xi>0$ and minus for $\xi<0$. We omit the channel index in $\xi$. Transformations \eqref{pozor0}-\eqref{pozor2} map the conformal boundary to the points $0$, $-16$ and $-1728$ in the scalar, shear, and sound channel, respectively.
In all three channels the horizon corresponds to $\xi=\infty$.
The transformed equations of motion have the form \eqref{forQM}, with the identifications $\hbar=\alpha^{-{1\over 4}}$ and 
  {\small{\begin{equation}\label{pot}
        \left\{V_{\rm scalar},V_{\rm shear},V_{\rm sound}\right\}=\left\{ \frac{4-\xi}{16\,\xi^{3\over2}},-\frac{\pm1}{64 \left(4\pm\abs{\xi}^\frac12\right)^{3\over2} \abs{\xi}^\frac12},-\frac{8+\xi^\frac13}{144 \left(12+\xi^\frac13\right)^{3\over2} \xi ^{4\over3}} \right\}\ ,
    \end{equation}}}where in $V_{\rm shear}$ we use the plus signs for $\xi>0$ and the minus ones for $\xi<0$. 
In all channels the potential forms a barrier ($V>0$) in the near-boundary region:
     \begin{align}
    \text{Scalar}&:\quad \xi\in(0,4) \ , \label{neg1}\\
    \text{Shear}&:\quad \xi\in(-16,0) \ , \label{neg2}\\
    \text{Sound}&:\quad \xi\in(-1728,-512) \ .\label{neg3}
    \end{align}
The potential becomes negative for large $\xi$. In addition, in the sound channel we find a singularity in the classically allowed region at $\xi=0$.

One may now follow the standard WKB analysis. Considering the ansatz of the form
    \begin{equation}
        Z(\xi)=e^{i\alpha^{-\frac14}(W_0(\xi)+\alpha^{-\frac14} W_1(\xi)+\ldots)} 
    \end{equation}
and plugging this into the transformed equations, one determines the functions $W_i(\xi)$. The leading term is
    \begin{equation}\label{lead}
        W_0(\xi)=\pm\int^\xi\sqrt{-V(\xi')}\dd \xi'\ .
    \end{equation}
In the classically forbidden region we pick the sign corresponding to the decaying exponential, while deep in the bulk we select the oscillating solution that satisfies the incoming-wave condition near the horizon $\xi=\infty$.
Validity of WKB is restricted to the region where
    \begin{equation}
\left|\dv{\xi}\sqrt{V(\xi)}\right|\ll\alpha^{1\over 4}\abs{V(\xi)}\ .
    \end{equation}
One cannot use the WKB ansatz close to the boundary and around the turning point -- in these regions one has to solve the  equations of motion and connect solutions inside and outside the barrier.
 
The imaginary part of the correlator, using \eqref{eq.origG},  can be expressed as the following tunnelling probability\footnote{See also Appendix B in \cite{Son:2002sd} for a related discussion.} 
    \begin{equation}\label{imgr}
        \Im\ff\sim\Im G_R\sim \exp\left[-2\alpha^{\frac14}\int_{\text{barrier}}\!\!\!\!\sqrt{V(\xi')}\,\dd \xi'\right]\ .
    \end{equation}
In all three channels, the imaginary part of the correlator decays exponentially at the same rate\footnote{Note the overall prefactors depend on the channels, as is apparent already from Fig.\ \ref{fig.imAll}. We verified that the WKB results agree with the numerics.} 
    \begin{equation}\label{wkbres}
        \Im\mathfrak{f}\sim\exp\left[-2\alpha^{\frac14}\int_0^2\sqrt{\frac{4-y^2}{4y}}\dd y\right]=\exp\left[-\frac{\sqrt{\frac{\pi }{2}} \Gamma \left(\frac{1}{4}\right)}{\Gamma \left(\frac{7}{4}\right)}\alpha^{1\over 4}\right]\approx e^{-4.94~ \alpha^{1\over 4}}\ .
    \end{equation} 
It would be interesting to see if this  behavior holds more generally.
Note that the exponent in the  retarded thermal correlator of a scalar field in the large momentum  limit was computed in \cite{Son:2002sd,Dodelson:2022yvn}.   
Although the near-lightcone limit we consider here is different, the power of momenta in the exponent is the same, \textit{i.e}.\ $\alpha^{1/4}\sim q$, while  the multiplicative constants in the exponent differ. 


\section{Discussion}\label{sec.discussion} 

In this paper we point out that the near-lightcone thermal correlators of stress tensors in holographic  Einstein-Gauss-Bonnet gravity take the same form as those in Einstein gravity.\footnote{This work focuses on $d=4$ and we expect similar results in other $d>2$ dimensions.}   More precisely, we observe that the thermal two-point correlators of stress tensors are rather constrained in the near-lightcone limit: they are given by three 
universal functions ($\ff_{\rm scalar},\ \ff_{\rm shear},\ \ff_{\rm sound}$) whose arguments involve the combination of $\alpha \sim q^+ (q^-)^3/T^4$  and 
the three coefficients  $\aa, \bb, \cc$ which determine the stress-tensor three-point functions.
The correlator in a given channel takes the vacuum form when the corresponding ANEC is saturated,
as already noticed in \cite{Huang:2022vet}.

The correlators  admit a perturbative expansion in powers of $1/\alpha$.
This is essentially the OPE  combined with the near-lightcone limit, where only the leading-twist multi-stress tensors contribute.
One can read off the OPE coefficients of the two stress tensors and multi-stress tensors.
The momentum space approach might be  more convenient than the one which employs the near-boundary ansatz in position space and substitutes it into the equations of motion, $e.g.$,  \cite{Karlsson:2022osn, Huang:2022vet}.
Note that the power series in momentum space seems to have zero radius of convergence, $i.e.$, it's an asymptotic series.\footnote{This was recently discussed in, $e.g.$, \ \cite{Caron-Huot:2009ypo,Iliesiu:2018fao,Manenti:2019wxs,Dodelson:2023vrw}.}

Depending on whether we want to compute retarded or advanced correlators, we need to impose appropriate boundary conditions at the horizon (which in our variables corresponds to the behavior at large $x$). 
Perturbatively, the correlator is completely determined by the OPE (as we explain in Section \ref{sec:Momentum} and Appendix \ref{ap.pertpos}).
However the boundary conditions at the horizon affect the solution non-perturbatively in $\alpha$.
This is because a general solution decays exponentially under the barrier, as discussed in Section \ref{sec.wkbim}.
It would be interesting to understand the significance of such non-perturbative terms.\footnote{In two spacetime dimensions, similar terms 
appear after the Fourier transform of the HHLL Virasoro vacuum block, see, $e.g.$, \cite{Son:2002sd,Manenti:2019wxs}.
(See also \cite{Haehl:2018izb,Datta:2019jeo,Jensen:2019cmr,Ramirez:2020qer} for explicit expressions of  retarded $TT$ correlators.) In four-dimensional holographic CFTs on the sphere, the spectrum of quasinormal modes contains contributions that are non-perturbative in spin \cite{Dodelson:2022eiz} (see also \cite{Festuccia:2008zx}). It would be interesting to understand to what extent the non-perturbative terms are determined by the OPE in $d=4$.}

In the near-lightcone limit, because of the universality of $TT$-correlators discussed above,
we can simply focus on the analysis based on pure Einstein gravity. 
Note that for the transverse polarization of the stress tensor, the equation of motion for the metric fluctuation is the same as that for 
a minimally coupled scalar.
Hence the two point functions must be identical.
Naively one may find this surprising, given that the OPE coefficients for a scalar contain poles at integer values of the scalar's conformal dimension \cite{Fitzpatrick:2019zqz}. This corresponds to the mixing with the double-trace operators and fixes the residue of the OPE coefficient of the latter, which ensures
the divergence is cancelled.
How does this work in the stress-tensor correlator case and how is this reflected in momentum space?
The answer to this question is that the logarithmic terms in position space which are produced by the cancellation of
the poles at $\Delta=4$ get Fourier transformed to the rational functions of momenta (or, in our limit, 
$\alpha$) in momentum space.
Indeed, in Appendix  \ref{apB}, we verify explicitly that the OPE coefficients of the two scalars and multi-stress tensors,
together with the thermal expectation values of the latter, multiplied by the corresponding conformal blocks in momentum space reproduce
the perturbative expansion of the transverse $TT$ correlator.

It is useful to examine the large-$N$ counting 
of  thermal  $TT$ correlators (where $N\sim \sqrt{C_T}$). Consider a finite temperature connected $TT$-correlator on a sphere 
above the confinement-deconfinement phase transition.
The disconnected component scales like $N^4$ and this behavior is entirely due to the double-stress tensor operators $[T_{\mu\nu}]^2$, as explained
in \cite{Karlsson:2022osn}. 
Indeed, the MFT OPE coefficients $\lambda_{T_{\mu\nu} T_{\alpha\beta} [T_{\mu\nu}]^2} \sim 1$ while $\langle [T_{\mu\nu}]^2 \rangle \sim N^4$.
The subleading corrections to the OPE coefficients and to the anomalous dimensions of $[T_{\mu\nu}]^2$ contribute to the connected correlator  \cite{Karlsson:2022osn}.
It is easy to extract the large-$N$ behavior of the OPE coefficients with the $k$-stress tensors and convince oneself that they all contribute to 
the connected correlator with the expected $N^2$ scaling. 

On the other hand, in the low-temperature phase $\langle [T_{\mu\nu}]^k \rangle $ scales like $N^0$, while the leading large-$N$ behavior of the $TT$ correlator scales like $ N^2$.
In holographic theories such correlators are simply given by the sum of the vacuum correlators over the thermal images.
One can immediately see how this is reproduced by multi-stress tensors.
The only contributions that survive in addition to the identity are the double-stress tensors.

There are various extensions to consider. Let us mention a few of them: (i) Extend the analysis of near-lightcone correlators to the charged black hole case. (ii) Understand more precisely how non-universal coefficients affect the thermal $TT$ correlators when moving slightly away from the lightcone limit. (iii) It may be useful to further study the thermal $TT$ correlators using CFT techniques developed in, $e.g.$, 
\cite{Caron-Huot:2017vep, Simmons-Duffin:2017nub, Iliesiu:2018fao, Gobeil:2018fzy, Petkou:2018ynm,Karlsson:2019dbd, Manenti:2019wxs,Delacretaz:2020nit,Alday:2020eua,Rodriguez-Gomez:2021pfh,Engelsoy:2021fbk,Parisini:2022wkb,Dodelson:2022yvn,Bhatta:2022wga,Avdoshkin:2022xuw,Fortin:2023czq,Dodelson:2023vrw}. 
(iv) It would also be interesting to see if anything useful can be said about the regime $|k|\gg|\omega|$ \cite{Banerjee:2019kjh,Caron-Huot:2022akb}. 

In this work, we speculate that the universality observed in the holographic Einstein-Gauss-Bonnet theory remains valid  in more general holographic theories.  It would be interesting to study the near-lightcone $TT$ correlators using different gravity models  to see if this universality persists. On the other hand, understanding this directly from the CFT point of view would be a more ambitious but very interesting goal. In this spirit a possible step is to study the lightcone limit of heavy-heavy-light-light correlators, with the light operators being stress tensors, from the bootstrap point of view. This would extend recent progress for scalar correlators in e.g.\ \cite{Karlsson:2019dbd,Li:2019zba} and related works. In the latter, the Lorentzian inversion formula was used to get the OPE data for multi-stress tensors in the scalar case and it would be interesting to generalize this to stress tensor correlators, or spinning correlators more generally. A CFT bootstrap approach would likely shed light on the regime of universality beyond the cases explored in this paper and is therefore of great interest.

\bigskip

\section*{Acknowledgments}

We would like to thank M. Dodelson, L. Iliesiu, D. Jafferis, M. Kulaxizi, H. Liu, V. Niarchos and S. Zhiboedov for useful discussions.
CE  was supported in part by the Irish Research Council Government of Ireland Postgraduate Fellowship  under project award number GOIPG/2022/887.
KWH was supported in part by the Irish Research Council Government of Ireland Postdoctoral Fellowship under project award number GOIPD/2022/288. 
RK was  supported by the European Research Council (ERC) under the European Union’s Horizon 2020 research and innovation programme (grant agreement number 949077).  
AP  and SV were supported in part by the  Irish Research Council Consolidator Award No. IRCLA/2017/82.
The research of AP was also supported in part by the Science Foundation Ireland and by NSF under Grant No. NSF PHY-1748958.
AP thanks KITP Santa Barbara and Harvard University, where part of this work was completed,  for hospitality.


\appendix

\section{Position-Space Analysis}\label{ap.pertpos}

To have a consistency check on the momentum-space results, here we discuss perturbative solutions of the reduced equations of motion in position space, based on the  computation performed in \cite{Karlsson:2022osn,Huang:2022vet}. In position space, we can systematically calculate the 
near-lightcone correlators 
order by order in a $\mu$ expansion where $\mu=(\pi/\beta)^4$.  

    \subsection{OPE in position space from holography}

Let the five-dimensional bulk coordinates\footnote{In this position-space calculation we adopt the Euclidean signature, as was done in \cite{Karlsson:2022osn,Huang:2022vet}.} be $(r,t_E,x,y,z)$ and assume the metric fluctuations do not depend $x$ and $y$.
We consider a near-boundary, OPE:
    \begin{equation}\label{eq.opel}
        r\rightarrow\infty\quad\text{with}\quad rt_E,\,rz\,\,\text{fixed} \ .
    \end{equation}
 Defining $v=\frac{z}{r}$ and $w^2=1+r^2t_E^2+r^2z^2$, the bulk limit that isolates the near-lightcone correlator contribution is \cite{Fitzpatrick:2019zqz, Huang:2022vet}
    \begin{equation}\label{eq.lcl}
        r\rightarrow\infty\quad\text{with}\quad v\,\,\text{fixed} \ .
    \end{equation}
Performing this limit on the equations of motion, one gets the reduced equations of motion for the bulk-to-boundary propagators $\mathcal{Z}$ 
which can be solved by the ansatz:\footnote{We rewrite the ansatz so it looks different from the one in \cite{Huang:2022vet}.}
    \begin{align}
        \mathcal{Z}&=\mathcal{Z}^{AdS}\left(Q+\bar{Q}(\log r-\log w)\right),\label{eq.ourversionQtot}\\
    Q=\sum_{n=0}^\infty\sum_{m=-n}^n&a_{nm}v^{2n}w^{2m}\quad\text{and}\quad
    \bar{Q}= \sum_{n=0}^\infty\sum_{m=3}^nb_{nm}v^{2n}w^{2m}\label{Qdef}
    \end{align}
where $\mathcal{Z}^{AdS}$ is the bulk-to-boundary propagator in pure AdS, which in the scalar channel is $\frac{2r^2}{\pi w^6}$.\footnote{The full solution $Z$ is connected to $\mathcal{Z}$ by $Z(t_E,z,r)=\int\dd t_E^{\prime}\dd z'\mathcal{Z}(t_E-t_E^{\prime},z-z')\hat{Z}(t_E^{\prime},z')$, where $\hat{Z}$ is the boundary value of the invariant $Z$. For other channels and more details on the ansatz, see \cite{Karlsson:2022osn} and \cite{Huang:2022vet}}

Using the above ansatz and the scalar-channel reduced equation of motion (in the position space) obtained in \cite{Huang:2022vet}, one finds that the coefficients $a_{n,3}$ (for all $n\geq3$)  in \eqref{Qdef} are undetermined. 
This reflects the fact that, by performing the limit \eqref{eq.opel} one loses the information deep in the bulk. 
However, one can check that these undetermined coefficients are always suppressed in the lightcone limit.

We next compute the holographic stress-tensor correlators perturbatively in the $\mu$ expansion. 
In doing so, we note that in the lightcone limit ($i.e.$, $\xm\rightarrow0$ where $x^{\pm}= t\pm z=-it_E\pm z$) the correlator is fully determined by the coefficients $a_{nn}$.
The first few terms of the near-lightcone correlator $G_{\rm scalar}$ in position space are 
{\small
\allowdisplaybreaks
    \begin{equation}\label{eq.G3pertPosSpRES}
    \begin{aligned}
      \lim_{\xm \to 0}  G_{\rm scalar}=~ - {\pi C_T \over \xm^2} \Big(\frac{1}{5\xp^3\xm}
        &+\frac{1}{100} \mu\\
        &+\frac{\xm \xp^3}{1200} \mu^2\\
        &-\frac{21 \xm^2 \xp^6\log(-\xp\xm)}{286000} \mu^3\\
        &+\frac{71\xm^3 \xp^9\log(-\xp\xm)}{9792000} \mu^4\\
        &-\frac{2303\xm^4\xp^{12}\log(-\xp\xm)}{5684800000} \mu^5 \Big)+{\cal O} (\mu^6) \ .
    \end{aligned}
    \end{equation}}Using the same method, one can generalize the position-space computation to other two channels. However, due to the computational complexity in position space, in this paper we analyze the correlators in other channels in momentum space.

The correlator \eqref{eq.G3pertPosSpRES}  depends on the combination $\xm\xpp^3$, consistent with the Fourier transformed results using the variable $\alpha$ which we discuss next.

\subsection{Fourier transform to momentum space}

We here transform the position-space correlator \eqref{eq.G3pertPosSpRES}  to momentum space, where the conjugate variables to $(x^+,x^-)$ are $(q_+,q_-)=-\frac12(q^-,q^+)$. 
Fourier transform of the zeroth-order contribution diverges and thus needs to be regularized. 
We use the dimensional regularization, where instead of  $(\xm\xp)^{-3}$ we consider $(\xm\xp)^{-3-\epsilon}$ and then take $\epsilon \to 0$.  The result is 
    {\small\begin{equation}
        -\frac{\pi ^2 C_T (q^+q^-)^2}{320\,\epsilon}-\frac{1}{320} \pi ^2 C_T (q^+q^-)^2 \left(\log \left(-q^+q^-\right)+2 \gamma -3-\log (4)\right)+\mathcal{O}\left(\epsilon\right) \ ,
    \end{equation}}where $\gamma$ is the Euler's constant. Terms $\mathcal{O}(\epsilon)$ can be neglected, while the pole can be eliminated by counterterms. 
The regulator-independent (physical) $\log (-q^+q^-) $ term is
   {\small \begin{equation}
    -\frac{\pi C_T}{5\xm^3\xp^3}\longrightarrow-\frac{1}{320} \pi ^2 C_T (q^+q^-)^2\log (-q^+q^-).
    \end{equation}}We have verified that this result exactly matches the leading term in \eqref{eq.0corrG3} computed in momentum space.

The $\OO(\mu^1)$ and $\OO(\mu^2)$ terms can be directly Fourier transformed: 
    {\small\begin{align}
        -\frac{\pi C_T\mu}{100\xm^2}\longrightarrow-\frac{\pi ^2 C_T \mu  (q^++q^-)}{100 q^-}\approx-\frac{\pi ^2 C_T \mu  q^+}{100 q^-}\ , ~~
        -\frac{\pi C_T\mu^2\xp^3}{1200\xm}\longrightarrow\frac{2\pi^2 C_T \mu ^2}{25 (q^-)^4} \ .
    \end{align}}
\begin{center}
\begin{figure}[h]
\hspace*{-1.5cm}
	\includegraphics[scale=0.72]{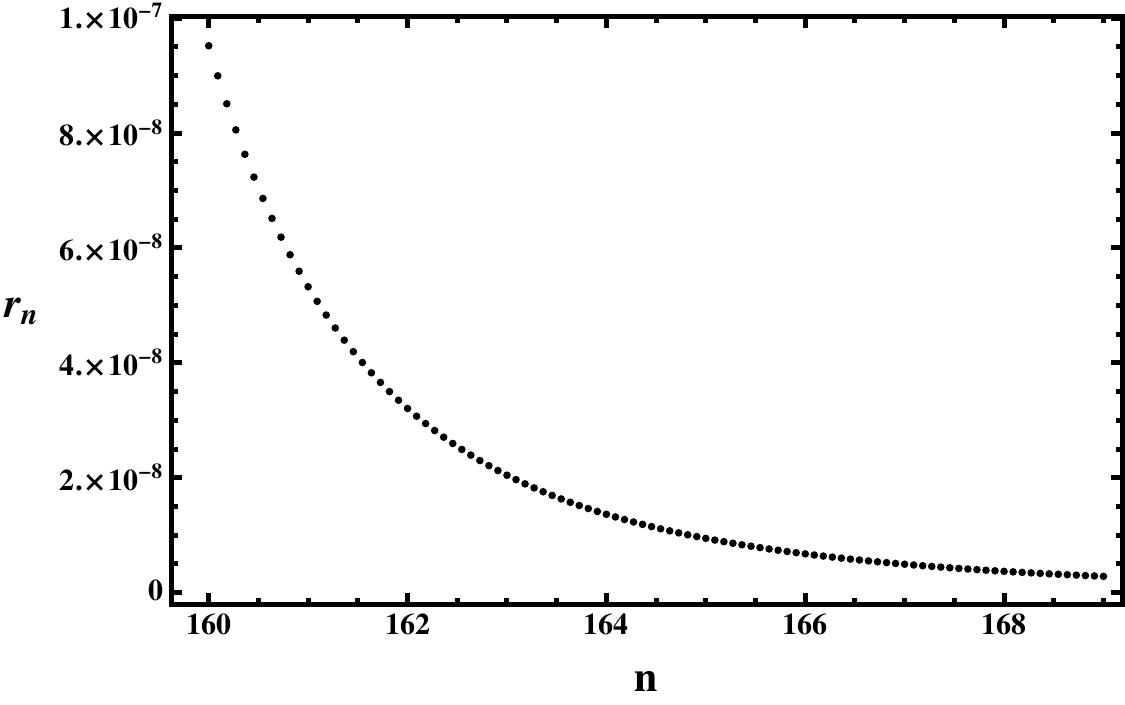}
    \centering
    \caption{Estimation of the radius of convergence in the scalar channel}
    \label{fig.rc}
\end{figure}
\end{center}
For the terms $\OO(\mu^n)$ with $n\geq3$, we use
    {\small\begin{equation}
    \log(-\xp\xm)\longrightarrow\frac{4\pi}{q^+q^-} \ .
    \end{equation}}Applying derivatives with respect to $\xp$ and $\xm$  gives   
    {\small\begin{equation}\label{eq.alllogft}
        \xpp^{-6}(\xp^3\xm)^n\log(-\xp\xm)\longrightarrow-\frac{3 \pi  16^{n+2} \Gamma (n+3) \Gamma (3 n+6) }{(n+1) q^+(q^-)^7\left(q^+(q^-)^3\right)^{n}} \ ,
    \end{equation}}which is valid for any $n\geq0$. Using this result, we can Fourier transform the log-terms appearing in \eqref{eq.G3pertPosSpRES}.
We obtain
{\small\begin{align}
G=G_0\Big( -\frac{1}{2}\log(-q^+q^-)  + \sum_{k= 1} c_{k}\alpha^{-k} \Big) \ , ~~~\alpha=-\mathfrak{q}^+(\mathfrak{q}^-)^3 
\end{align}}where $G_0$ and the first  several coefficients $c_{k}$ are
{\small\begin{align}
G^{\rm{scalar}}_0&= \frac{\pi^2C_T(q^+q^-)^2}{160} \ ,\,\,\,\,\, c^{\rm{scalar}}_{1}= \frac{1}{10} \ ,\,\,\,\,\,\, c^{\rm{scalar}}_{2}=  \frac{1}{20} \ , \,\,\,\,\,\,\,\, c^{\rm{scalar}}_{3}= \frac{378}{715}\ ,\ldots \label{scalarforri}\\ 
G^{\rm{shear}}_0&= - \frac{\pi^2C_T(q^+q^-)^2}{160} \ ,  \,c^{\rm{shear}}_{1}=- \frac{1}{40} \ , ~ c^{\rm{shear}}_{2}= -\frac{17}{560} \ , ~  c^{\rm{shear}}_{3}=-\frac{2241}{5720}\ ,\ldots \\
G^{\rm{sound}}_0&= - \frac{\pi^2C_T(q^+q^-)^2}{160} \ , \,c^{\rm{sound}}_{1}= \frac{1}{60} \ , \,\,\,\,\, c^{\rm{sound}}_{2}=\frac{11}{420} \ , \,\,\,\,\,\, c^{\rm{sound}}_{3}= \frac{2297}{6435}\ ,\ldots 
\end{align}}We further list the fourth-order correction in the scalar channel for comparison in Appendix \ref{apB}:
{\small\begin{align} \label{eq:scalarTQuar}
 c^{\rm{scalar}}_{4}=\frac{4473}{170} \ . 
\end{align}}Focusing on the scalar channel, in Fig.\ \ref{fig.rc} we also estimate the radius of convergence (see \eqref{rok})  by computing more higher-order terms in the expansion \eqref{scalarforri}. We find that $\lim_{n\to \infty} r_n \approx 0$.

\section{Momentum-Space Thermal Conformal Blocks} \label{apB}

In this appendix, we use the momentum-space conformal blocks to examine the scalar channel in Einstein gravity, whose EoM and action are equivalent to the ones for a massless scalar field. Note, however, that this equivalence do not persist when considering higher-derivative terms, such as Einstein-Gauss-Bonnet gravity.\footnote{More precisely, while the OPE coefficients for minimal-twist stress tensors in the minimally coupled scalar case do not depend on higher-derivative terms such as the Gauss-Bonnet coupling \cite{Fitzpatrick:2019zqz} except through the temperature, the scalar wave equation in Einstein-Gauss-Bonnet gravity is different from the scalar-channel EoM of metric perturbations.}  

The thermal conformal blocks in momentum space were computed in \cite{Manenti:2019wxs}. 
Expanded in thermal conformal blocks, the scalar correlator can be written as
{\small\begin{align}
		\langle \cal{O}\cal{O}\rangle_\beta(\omega_n,|\mathbf{q}|=q) = \sum_{\cal{O}_{\Delta,J}} a_{\cal{O}_{\Delta,J}}G_{\Delta,J}^{\Delta_\cal{O}}(\omega_n,q) \ , ~~
		a_{\cal{O}_{\Delta,J}} = \frac{\lambda_{\cal{O}\cal{O}\cal{O}_{\Delta,J}}b_{\cal{O}_{\Delta,J}}}{c_{\cal{O}_{\Delta,J}}\beta^\Delta}\frac{J!}{2^J(\frac{d-2}{2})_J} 
\end{align}}where $\omega_n=\frac{2\pi n}{\beta}$ is the Matsubara frequency and $a_{\cal{O}_{\Delta,J}}$ are thermal coefficients. (The explicit form of  $G_{\Delta,J}^{\Delta_\cal{O}}(\omega_n,q)$ is given by Eq. (2.11) in \cite{Manenti:2019wxs}.)  Minimal-twist operators are the ones that dominate 
in the near-lightcone limit that we are interested in.

The relevant coefficients for the near-lightcone correlator up to $\cal{O}(\beta^{-16})$, where the contributing operators are $[T^k]_{J}$ with dimension $4k$ and spin $J=2k$ with $k=1,\ldots 4$, are 
{\allowdisplaybreaks
	{\small\begin{align}\label{eq:AT}
			a_1 &= \frac{3C_T}{10} \ , \\
			a_{T} &= \frac{3C_T}{10}\frac{\Delta_{\cal O}}{120}\left(\frac{\pi}{\beta}\right)^4 \ ,\\
			a_{[T^2]_{4}} &= \frac{3C_T}{10}\frac{\Delta_{\cal O}  \left(7 \Delta_{\cal O} ^2+6 \Delta_{\cal O} +4\right) }{201600 (\Delta_{\cal O} -2)}\left(\frac{\pi}{\beta}\right)^8 \ ,\\
			a_{[T^3]_{6}} &=  \frac{3C_T}{10}\frac{\Delta_{\cal O}  \left(1001 \Delta_{\cal O} ^4+3575 \Delta_{\cal O} ^3+7310 \Delta_{\cal O} ^2+7500 \Delta_{\cal O} +3024\right)}{10378368000 (\Delta_{\cal O} -3) (\Delta_{\cal O} -2)}\left(\frac{\pi}{\beta}\right)^{12}\ , \\
			a_{[T^4]_{8}}&=  \frac{3C_T}{10}\frac{\Delta_{\cal O}\left(119119 \Delta_{\cal O}  ^6+969969 \Delta_{\cal O}  ^5+4184550 \Delta_{\cal O}  ^4 \right) }{592812380160000
				(\Delta_{\cal O}  -4) (\Delta_{\cal O}  -3) (\Delta_{\cal O}  -2)}\left(\frac{\pi}{\beta}\right)^{16}\nonumber\\ \label{eq:AT4}
			&~~+\frac{3C_T}{10}\frac{\Delta_{\cal O}\left(10867340 \Delta_{\cal O}^3+16958856 \Delta_{\cal O} ^2+14428176 \Delta_{\cal O}+5009760   \right)}{592812380160000
				(\Delta_{\cal O}  -4) (\Delta_{\cal O}  -3) (\Delta_{\cal O}  -2)}\left(\frac{\pi}{\beta}\right)^{16}.
\end{align}}}We have used $\mu 
=(\pi /\beta)^4$ and normalized the correlator to agree with the stress-tensor correlator in the scalar channel, $\langle T_{xy}T_{xy}\rangle$.  
The stress-tensor coefficient is fixed by Ward identities and the stress-tensor one-point function, while the $[T^2]_4$ and $[T^3]_6$ were computed in \cite{Fitzpatrick:2019zqz,Karlsson:2019dbd}. Here we have further obtained the coefficient for the $[T^4]_8$ operator with dimension $16$ and spin $8$ based on the method of \cite{Karlsson:2019dbd}.  

Let us first discuss operators $[T^k]_{J=2k}$ with $k=0,1,2,3$.
For the identity contribution, the block has a simple pole at $\Delta_\cal{O}=4$ with a residue that is purely a contact term. 
Removing the contact term gives
\be 
\langle \cal{O}\cal{O}\rangle|_{1} = -\frac{\pi ^2C_T}{640} \left(q^2+\omega_n^2\right)^2 \log \left(q^2+\omega_n^2\right) \ .
\ee 
After Wick-rotating $\omega_n\to -i\omega$ and taking the lightcone limit we reproduce \eqref{eq.0corrG3}. 
Likewise, for the stress-tensor exchange we find 
\be 
\langle \cal{O}\cal{O}\rangle|_{T} = -\frac{C_T \pi^2}{200}\frac{q^-}{q^+}\left(\frac{\pi}{\beta}\right)^4 \ ,
\ee 
which is in agreement with \eqref{eq:TExch}, where we remind the reader that $G_{xy,xy}=\frac{1}{2}G_{\rm scalar}$.  Moreover, we have verified that the $[T^2]_4$ and $[T^3]_6$ contributions reproduce the coefficients listed in \eqref{scalarforri}.  

Something interesting happens when we consider $[T^k]_{2k}$ with $k\geq 4$. First notice that, as seen in \eqref{eq:AT4}, there is a pole at $\Delta_\cal{O}=4$.
Interpreted as a correlator of a scalar with dimension four, this is related to the operator mixing between the $[T^4]_8$ operator with dimension $16$ and spin $8$ and a double-trace operator of the schematic form $\cal{O}\partial_{(\mu}\partial_\nu\partial_\rho\partial_{\sigma)}\cal{O}$ with the same quantum numbers. However, the block with $\Delta=16$, $J=8$ has a zero at $\Delta_\cal{O}\to 4$:
{\small\begin{align}
		G_{16,8}^{\Delta_\cal{O}}(\omega_n,q)\propto\frac{(\Delta_{\cal O}-4) \left(q^8-36 q^6 \omega_n^2+126 q^4 \omega_n^4-84 q^2 \omega_n^6+9 \omega_n^8\right)}{\left(q^2+\omega_n^2\right)^{10}}+\cal{O}\big((\Delta_\cal{O}-4)^2\big) \ .
\end{align}}Thus, the pole in $a_{[T^4]_{8}}$ cancels with the zero of the blocks. By Wick-rotating and taking the lightcone limit, we obtain
\be 
\langle \cal{O}\cal{O}\rangle|_{[T^4]_{8}}=\frac{2290176 \pi ^2 C_T}{425 (q^-)^2(q^+)^{10}}\left(\frac{\pi}{\beta}\right)^{16} \ ,
\ee which is in agreement with \eqref{eq:scalarTQuar}.


\section{Equations of Motion for Einstein-Gauss-Bonnet Gravity}\label{ap.eomsGB} 

In Section \ref{sec:Momentum}, we showed that the equations of motion of metric fluctuations in Einstein-Gauss-Bonnet gravity in the limit \eqref{LCbulklimit} reduce to the equations of motion in Einstein gravity. Here we list the coefficients $A, B$ in the Einstein-Gauss-Bonnet  equations of motion using the notation adopted in this paper.\footnote{In Einstein-Gauss-Bonnet gravity, the equations of motion in terms of gauge-invariant variables were first derived in \cite{Buchel:2009sk}. Some simplified expressions can be found in,  $e.g.$, Appendix D of \cite{Grozdanov:2016fkt}.} 
\\\\
\noindent {\bf Scalar Channel:}
{\allowdisplaybreaks
{\small
\begin{align}
\label{sc1}
A_{\rm{scalar}}&=\frac{u}{u^2-1}  \left(\frac{1}{\left(\kappa ^2-1\right) \left(1-u^2\right)+1}+\frac{1}{U(u)}\right)-\frac{1}{u} \ , \\
\label{sc2}
B_{\rm{scalar}}&= \frac{1}{4 u (U(u)-1)} \Big(\frac{(\kappa -1) (\kappa +1)^2 \left(3 \left(\kappa ^2-1\right) u^2-\kappa ^2\right)}{U(u)^2}\qfr^2 +\frac{\left(\kappa ^2-1\right)^2 }{U(u)-1}\wfr^2 \Big)
\end{align}}}
where $U(u)= \sqrt{\kappa ^2-\kappa ^2 u^2+u^2}$. 

\vspace{0.5cm}

\noindent{\bf Shear Channel:}
{\allowdisplaybreaks
{\small
\begin{align}
\label{sh1}
A_{\rm{shear}}&=  {1\over {u (1-U(u)) U(u)^3 \big(\kappa ^2 (\kappa +1) q^2 (U(u)-1)-\left(\kappa ^2-1\right) \omega ^2 U(u)^2\big)}}\\
&~~~ \times \Big[2 \kappa ^4 (\kappa +1) \left(\frac{1}{2} \left(1-\kappa ^2\right) \left(u^2-1\right) (U(u)-2)+U(u)-1\right)\qfr^2  \nn\\
&~~~~ +\left(1-\kappa ^2\right)  \left(\kappa ^4+\left(1-\kappa ^2\right)^2 u^4-2 \left(1-\kappa ^2\right) u^2 \left(U(u)-\kappa ^2\right)-\kappa ^2 U(u)\right)  U(u)^2\wfr ^2 \Big]
\ , \nn\\
\label{sh2}
B_{\rm{shear}}&=  \frac{\kappa^2 (\kappa +1) (U(u)+1)}{4 u \left(u^2-1\right) U(u)^2} \qfr^2  +\frac{ U(u)^2+2 U(u)+1}{4 u \left(u^2-1\right)^2} \wfr^2
\end{align}}} 

\noindent{\bf Sound Channel:}
{\allowdisplaybreaks
{\small
\begin{align}
\label{so1}
A_{\rm{sound}}&={1\over {2 u D_1(u) (U(u)-1) U(u)^2}}\Big[3 D_1(u) U(u)^2 (U(u)-1)  + \Big(\left(\kappa ^2-1\right)^2 u^4 \left(-3 \kappa ^2+5 U(u)-7\right)\nn\\
&~~~~~~~~~~~ +\kappa ^2 \left(\kappa ^2-1\right) u^2 \left(18 \kappa ^2-13 U(u)+10\right)-15 \kappa ^4 \left(\kappa ^2-2 U(u)+1\right)\Big)\qfr^2\nn\\
&~~~~~~~~~~~ -3 (1-\kappa ) \left(\kappa ^2-\left(\kappa ^2-1\right) u^2\right) \big(5 \kappa ^2 (U(u)-1)-\left(\kappa ^2-1\right) u^2 (5 U(u)-7)\big) \wfr ^2 \Big]  \ ,\\ 
\label{so2}
B_{\rm{sound}}&=  \frac{\left(\kappa ^2-1\right)^2}{D_0(u)} \Big[
\left(\kappa ^2-1\right)^3 \qfr^2 u^6 \left(3 (\kappa -1) \wfr^2+\qfr^2\right)+12 (\kappa -1)^2 \kappa ^2 (\kappa +1) \qfr^2 u^5\nn\\
&~~~~~~-4 (\kappa -1) \kappa ^2 \qfr^2 u^3 \left(3 \kappa ^2-7 U(u)+4\right)+\left(\kappa ^2-1\right)^2 u^4 \Big(\qfr^4 \left(3 \kappa ^2 (U(u)-2)+U(u)\right)\nn\\
&~~~~~~+2 (\kappa -1) \qfr^2 \wfr^2 U(u)-3 (\kappa -1)^2 \wfr^4 U(u)\Big)-\kappa ^2 \left(\kappa ^2-1\right) u^2 \big(\qfr^4 \left(\kappa ^2+2 U(u)\right) \nn\\
&~~~~~~+(\kappa -1) \qfr^2 \wfr^2 \left(9 \kappa ^2-4 U(u)\right)-6 (\kappa -1)^2 \wfr^4 U(u)\big)-3 \kappa ^4 \big(\qfr^4 \left(\kappa ^2 (U(u)-2)+U(u)\right)\nn\\
&~~~~~~+2 (\kappa -1) \qfr^2 \wfr^2 \left(U(u)-\kappa ^2\right)+(\kappa -1)^2 \wfr^4 U(u)\big) \Big]
\end{align}}}
where
{\allowdisplaybreaks
{\small 
\begin{align}
D_0(u)&= 4 (\kappa -1) u (U(u)-1)^2 U(u)^3 D_1(u) \ , \\
D_1(u)& =  (\kappa ^2-1) u^2 \left(\qfr^2+3 (\kappa -1) \wfr^2\right)+3 \kappa ^2 \left( (U(u)-1)\qfr^2-(\kappa -1) \wfr ^2\right) 
\end{align}}}


\vspace{-0.5 cm}

\bibliographystyle{JHEP}
\bibliography{refs} 

\end{document}